\documentclass[11pt]{article}
\usepackage{latexsym,amssymb,amstext,amsmath}
\usepackage{hyperref}
\renewcommand{\baselinestretch}{1.2}
\allowdisplaybreaks
\def\Slash#1{\rlap{\hbox{$\mskip 3 mu /$}}#1}      
\newcommand{\ft}[2]{{\textstyle\frac{#1}{#2}}}

\begin{document}
%
\begin{titlepage}
\begin{flushright} \small
 ITP-UU-10/26 \\  Nikhef-2010-024
\end{flushright}
\bigskip

\begin{center}
 {\LARGE\bfseries  New supersymmetric higher-derivative
   couplings:\\[1.6ex]
 Full N=2 superspace does not count!}
\\[10mm]
\textbf{Bernard de Wit$^{a,b}$, Stefanos Katmadas$^a$ and Maaike van
  Zalk$^a$}\\[5mm]
\vskip 4mm
$^a${\em Institute for Theoretical Physics, Utrecht
  University,} \\
  {\em Leuvenlaan 4, 3584 CE Utrecht, The Netherlands}\\
$^b${\em Nikhef Theory Group, Science Park 105, 1098 XG Amsterdam, The
  Netherlands}\\[3mm]
{\tt B.deWit@uu.nl}\,,\; {\tt S.Katmadas@uu.nl}\,,\; {\tt
  M.vanZalk@uu.nl}
\end{center}

\vspace{3ex}

\begin{center}
{\bfseries Abstract}
\end{center}
\begin{quotation} \noindent An extended class of $N\!=\!2$ locally
  supersymmetric invariants with higher-derivative couplings based on
  full superspace integrals, is constructed. These invariants may
  depend on unrestricted chiral supermultiplets, on vector
  supermultiplets and on the Weyl supermultiplet. Supersymmetry is
  realized off-shell. A non-renormalization theorem is proven
  according to which none of these invariants can contribute to the
  entropy and electric charges of BPS black holes. Some of these
  invariants may be relevant for topological string deformations.
\end{quotation}

\vfill

\end{titlepage}

\section{Introduction}
\label{sec:introduction}
\setcounter{equation}{0}
Supersymmetric invariants with higher-derivative couplings play a role
in many applications. Here we will be dealing with $N\!=\!2$
supergravity, where the first higher-derivative couplings that were
considered involve the square of the Weyl tensor coupled to vector
supermultiplets \cite{Bergshoeff:1980is}. This particular class of
invariants is based on an integration over a chiral subspace of
$N\!=\!2$ superspace. It is relevant for the topological string
\cite{Antoniadis:1993ze,Bershadsky:1993cx}, and furthermore, it has
important implications for BPS black hole entropy \cite{Lopes
  Cardoso:1998wt}. Another class of invariants for vector multiplets
that involve terms quartic in the field strengths, was derived in
terms of $N=1$ superfields, both for the abelian
\cite{Henningson:1995eh} and for the non-abelian case
\cite{deWit:1996kc}. Unlike the previous class, this one is based on
an integral over full superspace. It yields important contributions to
the effective action of $N=2$ supersymmetric gauge theories (for some
additional references, see e.g.,
\cite{Dine:1997nq,Buchbinder:1999jn,Banin:2002mf,Argyres:2003tg}). A
related class of locally supersymmetric higher-derivative couplings was
considered in \cite{Morales:1996bp, Antoniadis:2010iq}. Those couplings
which involve both the Weyl tensor and higher-order coupling of the
vector field strengths, were conjectured to describe certain
deformations of the topological string partition function.

This paper deals with an explicit construction of this rather large
class of invariant couplings based on full superspace integrals. They
are coupled to conformal supergravity and are realized off-shell. This
feature greatly facilitates their construction, which is based on
previous work on $N\!=\!2$ supergravity (in particular, on
\cite{deRoo:1980mm,deWit:1980tn}). The general procedure underlying
this construction will be presented, and, as an explicit example,
many of the bosonic terms of the supergravity-coupled invariants that
contain $F^4$-, $R^2F^2$-, and $R^4$-terms, will be discussed. Here
$F$ denotes the abelian vector multiplet field strengths and $R$ the
Riemann tensor.

One of the motivations for this work is to study the possible
contribution of these new couplings to the entropy and the electric
charges of BPS black holes. As it turns out we can derive a
`non-renormalization' theorem according to which these contributions
vanish. This result is not entirely unexpected, in view of the fact
that there was already a good agreement for the subleading
contributions to the BPS entropy obtained from microstate counting and
from supergravity, in which the new couplings had so far not been
incorporated. Hence the existence of the non-renormalization theorem
offers a partial explanation for this agreement.

This paper is organized as follows. Section
\ref{sec:chiral-anti-chiral} presents the superconformal
transformations of chiral supermultiplets in a conformal supergravity
background as well as a number of related issues. Section
\ref{sec:kinet-chir-mult} describes the general strategy for the
construction of the higher-derivative couplings, based on the use of
the so-called `kinetic supermultiplet', which can be constructed from
an anti-chiral supermultiplet of zero Weyl weight. The components of
this multiplet are given in considerable detail, fully taking into
account the presence of the superconformal background. The
construction of the bosonic terms of the higher-derivative couplings
is presented in section \ref{sec:inv-higher-der-coulings}, together
with explicit examples based on a class of Lagrangians that
involves terms such as $F^4$, $R^2F^2$ and $R^4$. A
non-renormalization theorem pertaining to the entropy and the electric
charges of BPS black holes is proven in section
\ref{sec:non-renorm-theor-BPS}. Some concluding remarks are presented
in section \ref{sec:conclusions}.

In view of future applications and for the convenience of the reader
we have added four appendices, \ref{App:SC},
\ref{App:cov-supercon-boosts}, \ref{App:chiral-multiplets}, and
\ref{App:reduced-multiplets}, containing basic results on the
superconformal multiplet calculus. Many of these results have appeared
at various places in the literature, but we have updated them in
uniform notation. While some of them may not have been overly relevant
in the past, they are now required in the context of the new invariant
couplings.

\section{Chiral multiplets}
\label{sec:chiral-anti-chiral}
\setcounter{equation}{0}
Chiral superfields in flat $N=2$ superspace were first discussed in
\cite{Firth:1974st}. Subsequently they were derived in a conformal
supergravity background \cite{deRoo:1980mm,deWit:1980tn}. The latter
result was formulated in components and the same approach is followed
in this paper, although it is convenient to make use of superfield
notions at the same time. Chiral multiplets are complex and $N=2$
superspace is based on four chiral and four anti-chiral anticommuting
coordinates, $\theta^i$ and $\theta_i$, so that a scalar chiral
multiplet contains two times $2^4$ field components. These multiplets
carry a Weyl weight $w$ and a chiral $\mathrm{U}(1)$ weight $c$, which
is opposite to the Weyl weight, i.e. $c=-w$. The weights indicate
how the lowest-$\theta$ component of the superfield scales under Weyl
and chiral $\mathrm{U}(1)$ transformations. Anti-chiral multiplets can
be obtained from chiral ones by complex conjugation, so that
anti-chiral multiplets will have equal Weyl and chiral weights, hence
$w=c$.

The components of a generic scalar chiral multiplet are a complex
scalar $A$, a Majorana doublet spinor $\Psi_i$, a complex symmetric
scalar $B_{ij}$, an anti-selfdual tensor $F_{ab}^-$, a Majorana
doublet spinor $\Lambda_i$, and a complex scalar $C$. The assignment
of their Weyl and chiral weights is shown in
table~\ref{table:chiral}. The Q- and S-supersymmetry transformations
for a scalar chiral multiplet of weight $w$, are as
follows\footnote{
  \label{ftn:notation}
  Observe that \cite{deRoo:1980mm,deWit:1980tn} employ different
  conventions, in particular for (anti)symmetrization. Here
  (anti)symmetrization is always applied with unit strength.
}, 
\begin{align}
  \label{eq:conformal-chiral}
  \delta A =&\,\bar\epsilon^i\Psi_i\,, \nonumber\\[.2ex]
  \delta \Psi_i =&\,2\,\Slash{D} A\epsilon_i + B_{ij}\,\epsilon^j +
  \tfrac12   \gamma^{ab} F_{ab}^- \,\varepsilon_{ij} \epsilon^j + 2\,w
  A\,\eta_i\,,  \nonumber\\[.2ex]
  \delta B_{ij} =&\,2\,\bar\epsilon_{(i} \Slash{D} \Psi_{j)} -2\,
  \bar\epsilon^k \Lambda_{(i} \,\varepsilon_{j)k} + 2(1-w)\,\bar\eta_{(i}
  \Psi_{j)} \,, \nonumber\\[.2ex]
  \delta F_{ab}^- =&\,\tfrac12
  \varepsilon^{ij}\,\bar\epsilon_i\Slash{D}\gamma_{ab} \Psi_j+
  \tfrac12 \bar\epsilon^i\gamma_{ab}\Lambda_i
  -\tfrac12(1+w)\,\varepsilon^{ij} \bar\eta_i\gamma_{ab} \Psi_j \,,
  \nonumber\\[.2ex]
  \delta \Lambda_i =&\,-\tfrac12\gamma^{ab}\Slash{D}F_{ab}^-
   \epsilon_i  -\Slash{D}B_{ij}\varepsilon^{jk} \epsilon_k +
  C\varepsilon_{ij}\,\epsilon^j
  +\tfrac14\big(\Slash{D}A\,\gamma^{ab}T_{abij}
  +w\,A\,\Slash{D}\gamma^{ab} T_{abij}\big)\varepsilon^{jk}\epsilon_k
  \nonumber\\
  &\, -3\, \gamma_a\varepsilon^{jk}
  \epsilon_k\, \bar \chi_{[i} \gamma^a\Psi_{j]} -(1+w)\,B_{ij}
  \varepsilon^{jk}\,\eta_k + \tfrac12 (1-w)\,\gamma^{ab}\, F_{ab}^-
    \eta_i \,, \nonumber\\[.2ex]
    \delta C =&\,-2\,\varepsilon^{ij} \bar\epsilon_i\Slash{D}\Lambda_j
  -6\, \bar\epsilon_i\chi_j\;\varepsilon^{ik}
    \varepsilon^{jl} B_{kl}   \nonumber\\
  &\, -\tfrac14\varepsilon^{ij}\varepsilon^{kl} \big((w-1)
  \,\bar\epsilon_i \gamma^{ab} {\Slash{D}} T_{abjk}
    \Psi_l + \bar\epsilon_i\gamma^{ab}
    T_{abjk} \Slash{D} \Psi_l \big) + 2\,w \varepsilon^{ij}
    \bar\eta_i\Lambda_j \,.
\end{align}
The spinors $\epsilon^i$ and $\eta_i$ are the positive chirality
spinorial parameters associated with Q- and S-supersymmetry. The
corresponding negative chirality parameters are denoted by
$\epsilon_i$ and $\eta^i$. We note that hermitian conjugation is
always accompanied by raising or lowering of the $\mathrm{SU}(2)$
indices.
%
\begin{table}[t]
\begin{center}
\begin{tabular*}{10.8cm}{@{\extracolsep{\fill}}|c||cccccc| }
\hline
 & & \multicolumn{4}{c}{Chiral multiplet} & \\  \hline \hline
 field & $A$ & $\Psi_i$ & $B_{ij}$ & $F_{ab}^-$& $\Lambda_i$ & $C$ \\[.5mm] \hline
$w$  & $w$ & $w+\tfrac12$ & $w+1$ & $w+1$ & $w+\tfrac32$ &$w+2$
\\[.5mm] \hline
$c$  & $-w$ & $-w+\tfrac12$ & $-w+1$ & $-w+1$ & $-w+\tfrac32$ &$-w+2$
\\[.5mm] \hline
$\gamma_5$   & & $+$ & &   & $+$ & \\ \hline
\end{tabular*}
\vskip 2mm
\renewcommand{\baselinestretch}{1}
\parbox[c]{10.8cm}{\caption{\label{table:chiral}{\footnotesize
Weyl and chiral weights ($w$ and $c$) and fermion
chirality $(\gamma_5)$ of the chiral multiplet component fields.}}}
\end{center}
\end{table}

The transformation rules (\ref{eq:conformal-chiral}) are linear in the
chiral multiplet fields, and contain also other fields associated with
the conformal supergravity background, such as the self-dual tensor
field $T_{abij}$ and the spinor $\chi^i$. Other conformal supergravity
fields are contained in the superconformal derivatives $D_\mu$. The
superconformal multiplet of fields is described in more detail in
appendix \ref{App:SC}.

Products of chiral superfields constitute again a chiral superfield,
whose Weyl weight is equal to the sum of the Weyl weights of the
separate multiplets. Also functions of chiral superfields may describe
chiral superfields, assuming that they can be assigned a proper Weyl
weight. For instance, homogeneous functions of chiral superfields of
the same Weyl weight $w$ define a chiral supermultiplet whose Weyl
weight equals the product of $w$ times the degree of homogeneity. The
relevant formulae are presented in appendix
\ref{App:chiral-multiplets}.

Chiral multiplets of $w=1$ are special, because they are reducible
\cite{Firth:1974st,deRoo:1980mm}. Some details about these multiplets
are given in appendix \ref{App:reduced-multiplets}. For a scalar
chiral multiplet with $w=1$ the tensor $F_{ab}^-+F_{ab}^+$ is subject
to a Bianchi identity, which can be solved in terms of a vector gauge
field. The reduced scalar chiral multiplet thus describes the
covariant fields and field strength of a {\it vector multiplet}, which
encompasses $8+8$ bosonic and fermionic
components. Table~\ref{table:vector} summarizes the Weyl and chiral
weights of the various fields belonging to the vector multiplet: a
complex scalar $X$, a Majorana doublet spinor $\Omega_i$, a vector
gauge field $W_\mu$, and a triplet of auxiliary fields $Y_{ij}$. There
also exists an anti-selfdual tensor version of the chiral multiplet
with $w=1$ that is reducible. This multiplet is the so-called {\it
  Weyl supermultiplet}, which contains all the covariant fields and
curvatures of $N=2$ conformal supergravity. It contains $24+24$
bosonic and fermionic degrees of freedom. Both vector supermultiplets
and the Weyl multiplet play a central role in this paper.

\begin{table}[t]
\begin{center}
\begin{tabular*}{6.5cm}{@{\extracolsep{\fill}}|c||cccc| }
\hline
 & & \multicolumn{2}{c}{Vector multiplet} & \\  \hline \hline
 field & $X $ & $\Omega_i$ & $W_\mu$ & $Y_{ij}$   \\[.5mm] \hline
$w$  & $1$ & $ \tfrac{3}{2} $ & 0 &  2   \\[.5mm] \hline
$c$  & $-1$ & $- \tfrac12$ & 0 &  0   \\[.5mm] \hline
$\gamma_5$   &  & $ + $ &   &       \\ \hline
\end{tabular*}
\vskip 2mm
\renewcommand{\baselinestretch}{1}
\parbox[c]{6.5cm}{\caption{\label{table:vector}{\footnotesize
Weyl and chiral weights ($w$ and $c$) and fermion
chirality $(\gamma_5)$ of the vector multiplet component fields.}}}
\end{center}
\end{table}

Another special chiral multiplet is the so-called `kinetic' multiplet,
which has Weyl weight $w=2$. This multiplet is constructed from an
anti-chiral multiplet with $w=0$. It will be discussed in detail in
the next section.

Finally, scalar chiral multiplets with $w=2$ lead to superconformal
actions when including a conformal supergravity background. Their
highest component $C$ has Weyl weight 4, and chiral weight 0. To
define an action that is invariant under local superconformal
transformations one makes use of a density formula,
\begin{align}
  \label{eq:chiral-density}
  e^{-1}\mathcal{L} =&\, C - \varepsilon^{ij}\, \bar\psi_{\mu i} \gamma^\mu
  \Lambda_j-\tfrac18\bar \psi_{\mu i} T_{ab\,jk}\gamma^{ab}\gamma^\mu
  \Psi_l \,\varepsilon^{ij}\varepsilon^{kl}
   -\tfrac1{16}A( T_{ab\,ij} \varepsilon^{ij})^2 \nonumber\\
  &\,
  -\tfrac12\bar\psi_{\mu i}\gamma^{\mu\nu}\psi_{\nu j}\,
  B_{kl}\,\,\varepsilon^{ik}\varepsilon^{jl}
    + \varepsilon^{ij} \bar \psi_{\mu i}\psi_{\nu j}(F^{-\mu\nu}
    -\tfrac12 A\, T^{\mu\nu}{}_{kl}\,\varepsilon^{kl} )\nonumber\\
  &\,
  -\tfrac12 \varepsilon^{ij}\varepsilon^{kl} e^{-1}
  \varepsilon^{\mu\nu\rho\sigma} \bar\psi_{\mu i}\psi_{\nu j}
  (\bar\psi_{\rho k}\gamma_\sigma\Psi_{l} +\bar\psi_{\rho k}
  \psi_{\sigma j}\, A)\,.
\end{align}

\section{The kinetic chiral multiplet}
\label{sec:kinet-chir-mult}
\setcounter{equation}{0}
The term `kinetic' multiplet was first used in the context of the
$N=1$ tensor calculus \cite{Ferrara:1978jt}, because this is the
chiral multiplet that enables the construction of the kinetic terms,
conventionally described by a real superspace integral, in terms of a
chiral superspace integral. In flat $N=1$ superspace, this
construction is simply effected by the conversion,
\begin{equation}
  \label{eq:real-chiral-superspace}
  \int \mathrm{d}^2\theta \;\mathrm{d}^2\bar \theta\; \Phi\,\bar\Phi^\prime
  \approx
  \int \mathrm{d}^2\theta \; \Phi\, \mathbb{T}(\bar\Phi^\prime) \,,
\end{equation}
up to space-time boundary terms. Here $\Phi$ and $\Phi^\prime$ are two
chiral superfields and $\bar\Phi^\prime$ is the anti-chiral field
obtained from $\Phi^\prime$ by complex conjugation. The kinetic
multiplet equals $\mathbb{T}(\bar\Phi^\prime)=\bar D^2 \bar
\Phi^\prime$, where $\bar D$ denotes the supercovariant $\bar
\theta$-derivative. Obviously the kinetic multiplet contains terms
linear and quadratic in space-time derivatives, so that, upon
identifying $\Phi$ and $\Phi^\prime$, the right-hand side of
\eqref{eq:real-chiral-superspace} does indeed give rise to the kinetic
terms of an $N=1$ chiral multiplet.

In \cite{deWit:1980tn} a corresponding kinetic multiplet was
identified for $N=2$ supersymmetry, which now involves four rather
than two covariant $\bar\theta$-derivatives,
i.e. $\mathbb{T}(\bar\Phi)\propto \bar D^4 \bar \Phi$. As a result,
$\mathbb{T}(\bar\Phi)$ contains now up to four space-time derivatives,
so that the expression
\begin{equation}
  \label{eq:real-chiral-n2}
  \int \;\mathrm{d}^4\theta\;\mathrm{d}^4\bar\theta \;\Phi\,\bar\Phi^\prime
  \approx
  \int \;\mathrm{d}^4\theta \,\Phi\,\mathbb{T}(\bar\Phi^\prime) \,,
\end{equation}
does not correspond to a kinetic term, but to a higher-order
derivative coupling.  Furthermore, for $N=2$ supersymmetry one has the
option of expressing the chiral multiplets in terms of (products of)
reduced chiral multiplets. In that case, expressions such as
\eqref{eq:real-chiral-n2} will correspond to higher-derivative
couplings of vector multiplets. Since we are considering the kinetic
multiplets in a conformal supergravity background, their Weyl weight
is relevant. Both in $N=1,2$ supergravity the kinetic multiplet
carries Weyl weight $w=2$. The conversion starts from a $w=1$ chiral
multiplet for $N=1$ and from a $w=0$ chiral multiplet for $N=2$
supersymmetry, respectively.

To demonstrate this in more detail, consider an anti-chiral $N=2$
supermultiplet in the presence of the superconformal background. Its
supersymmetry transformations follow from taking the complex conjugate
of \eqref{eq:conformal-chiral}. Precisely for $w=0$ we note that the
field $\bar C$ is invariant under S-supersymmetry and transforms under
Q-supersymmetry as the lowest component of a chiral supermultiplet
with $w=2$. This observation proves that we are dealing with a $w=2$
chiral supermultiplet, as is also confirmed by the weight assignments
specified in table \ref{table:chiral}. What remains is to identify the
various components of this multiplet in terms of the underlying $w=0$
multiplet. This can be done by applying successive Q-supersymmetry
transformations on $\bar C$, something that requires rather tedious
calculations in the presence of a superconformal background.

Denoting the components of $\mathbb{T}(\bar \Phi_{w=0})$ by
$(A,\Psi,B,F^-,\Lambda,C)\vert_{\mathbb{T}(\bar\Phi)}$, while
$(A,\Psi,B,F^-,\Lambda,C)$ will denote the components of the original
$w=0$ chiral multiplet, we have established the following relation,
\begin{align}
  \label{eq:T-components}
  A\vert_{\mathbb{T}(\bar\Phi)} =&\, \bar C \,, \nonumber\\[.3ex]
  \Psi_i\vert_{\mathbb{T}(\bar\Phi)}=&\,-2\,\varepsilon_{ij}
  \Slash{D}\Lambda^j
  -6\, \;\varepsilon_{ik} \varepsilon_{jl} \chi^j B^{kl}
  -\tfrac14\varepsilon_{ij}\varepsilon_{kl} \, \gamma^{ab} T_{ab}{}^{jk}
  \stackrel{\leftrightarrow}{\Slash{D}}\Psi^l \,,  \nonumber\\[.6ex]
  B_{ij}\vert_{\mathbb{T}(\bar\Phi)} =&\, - 2\,\varepsilon_{ik}\varepsilon_{jl}
  \big(\Box_\mathrm{c} + 3\,D\big) B^{kl}   -2\, F^+_{ab}\,
  R(\mathcal{V})^{ab\,k}{}_{i}\, \varepsilon_{jk} \nonumber\\
  &\, -6\,\varepsilon_{k(i}\,\bar\chi_{j)}\Lambda^k + 3\, \varepsilon_{ik}
  \varepsilon_{jl} \bar\Psi^{(k}\Slash{D}\chi^{l)} \,,
  \nonumber \\[.6ex]
  F_{ab}^-\vert_{\mathbb{T}(\bar\Phi)} =&\, - \big(\delta_a{}^{[c} \delta_b{}^{d]}
    -\tfrac12\varepsilon_{ab}{}^{cd}\big)\nonumber\\
    &\quad \times \big[4\, D_cD^e F^+_{ed} + (D^e\bar A
    \,D_cT_{de}{}^{ij}+D_c\bar A
   \,D^eT_{ed}{}^{ij})\varepsilon  _{ij} \big] \nonumber\\
  &\, +\Box_\mathrm{c} \bar A \,T_{ab}{}^{ij}\varepsilon_{ij}
    -R(\mathcal{V})^-{}_{\!\!ab}{}^i{}_k \,B^{jk} \,\varepsilon_{ij}
    +\tfrac1{8} T_{ab}{}^{ij} \,T_{cdij} F^{+cd} - \varepsilon_{kl}\,
    \bar\Psi^k\stackrel{\leftrightarrow}{\Slash{D}} R(Q)_{ab}{}^l
    \nonumber\\
   &\, -\tfrac94 \varepsilon_{ij} \,\bar\Psi^i\gamma^c\gamma_{ab}D_c
    \chi^j +  3\,\varepsilon_{ij} \bar\chi^i\gamma_{ab}
    \Slash{D}\Psi^j +\tfrac3{8}T_{ab}{}^{ij}\varepsilon_{ij}\,
    \bar\chi_k\Psi^k \,,
    \nonumber\\[.6ex]
    \Lambda_i\vert_{\mathbb{T}(\bar\Phi)} =&\, 2\,\Box_\mathrm{c}\Slash{D}
    \Psi^{j}\varepsilon_{ij}
    + \tfrac1{4}  \gamma^c\gamma_{ab} (2\, D_c
      T^{ab}{}_{ij}\,\Lambda^{j} + T^{ab}{}_{ij} \,D_c \Lambda^{j})
      \nonumber\\
   &\,
   - \tfrac1{2}\varepsilon_{ij} \big(R(\mathcal{V})_{ab}{}^j{}_k +
   2\mathrm{i} \, R(A)_{ab}\delta^j{}_k\big)\,\gamma^c\gamma^{ab} D_c\Psi^k
   \nonumber\\
   &\,
   +\tfrac12\,\varepsilon_{ij} \big( 3\, D_b D
   - 4\mathrm{i} D^a R(A)_{ab}
     +\tfrac{1}{4}
     T_{bc}{}^{ij}\stackrel{\leftrightarrow}{D_a} T^{ac}{}_{ij}
     \big)\,\gamma^b \Psi^{j}  \nonumber\\
   &\,
   -2\,F^{+ab}\, \Slash{D}R(Q)_{ab}{}_{i}
   +6\,\varepsilon_{ij}  D\, \Slash{D} \Psi^{j}  \nonumber\\
   &\,       + 3 \,\varepsilon_{ij}\,\big(\Slash{D}\chi_k\,B^{kj}
     +\Slash{D} \bar A\,\Slash{D}\chi^{j}  \big)  \nonumber\\
   &\,
   + \tfrac32\big( 2\,\Slash{D}B^{kj} \varepsilon_{ij}
    +\,  \Slash{D} F_{ab}^+ \gamma^{ab} \, \delta^k_{i}
    +\tfrac14  \varepsilon_{mn} T_{ab}{}^{mn}\,\gamma^{ab}
    \,\Slash{D}\bar A\, \delta_i{}^k\big)  \chi_k  \nonumber\\
   &\, +\tfrac94 \, (\bar\chi^l\gamma_a\chi_l)
   \,\varepsilon_{ij}\gamma^a \Psi^{j}
    - \tfrac92 \, (\bar\chi_i\gamma_a\chi^k)
    \,\varepsilon_{kl}\gamma^a \Psi^{l} \,,  \nonumber\\[.6ex]
  C\vert_{\mathbb{T}(\bar\Phi)}=&\,
  4(\Box_\mathrm{c} + 3\, D) \Box_\mathrm{c} \bar A -\tfrac12
  D_a\big(T^{ab}{}_{ij} \,T_{cb}{}^{ij}\big) \,D^c\bar A
   +\tfrac1{16} (T_{abij}\varepsilon^{ij})^2 \bar C
  \nonumber\\
  &\,
  + D_a\big(\varepsilon^{ij} D^a T_{bc ij}\,F^{+bc} +4
  \,\varepsilon^{ij} T^{ab}{}_{ij} \,D^cF^{+}_{cb} -
     T_{bc}{}^{ij}\, T^{ac}{}_{ij} \,D^b\bar A \big)  \nonumber\\
  &\,
  + \big( 6\,D_b D   - 8\mathrm{i}D^a R(A)_{ab} \big) \,D^b \bar A
  +\cdots\,,
\end{align}
where in the last expression we suppressed terms quadratic in the
covariant fermion fields. Obviously terms involving the fermionic
gauge fields, $\psi_\mu{}^i$ and $\phi_\mu{}^i$, are already contained
in the superconformal derivatives. Observe that the right-hand side of
these expressions is always linear in the conjugate components of the
$w=0$ chiral multiplet, i.e. in $(\bar A,\Psi^i,B^{ij}, F^+_{ab},
\Lambda^i, \bar C)$. As an extra test of the correctness of
(\ref{eq:T-components}) we verified that these expressions satisfy the
correct transformation behaviour under S-supersymmetry. This test
cannot be performed on the last component
$C\vert_{\mathbb{T}(\bar\Phi)}$, because we refrained from collecting
the fermionic contributions. As an extra check we have therefore verified
that the bosonic terms of $C\vert_{\mathbb{T}(\bar\Phi)}$ are
invariant under special conformal boosts.

The definition of the superconformal D'Alembertian $\Box_\mathrm{c}$,
defined by the contraction of two superconformal derivatives $D_a$, as
well as multiple superconformal derivatives in general, may require
further comment. Therefore we have presented some relevant material in
appendix \ref{App:cov-supercon-boosts}. Below we give the most
non-trivial transformation rules under special conformal boosts that
are needed in this paper,
\begin{align}
  \label{eq:K-var-A-Psi}
  \delta_\mathrm{K} \Box_\mathrm{c}\Box_\mathrm{c} A =&\,
  - 2\,\Lambda_{\mathrm{K}}^a  \big([D_a,D_b] D^b +D^b[D_a,D_b]
  \big)A \nonumber  \\
  =&\,
  \tfrac14 \Lambda_\mathrm{K}^a \,T_{ac}{}^{ij}\,T^{bc}{}_{ij}\,
  D_bA -3\,\Lambda_{\mathrm{K}}{}^a D\,D_aA
     -2\,\Lambda_{\mathrm{K}}{}^a \,D^b(\bar R(Q)_{ba} \Psi_i)\nonumber \\
   &\, -\tfrac34 \Lambda_{\mathrm{K}}{}^a \,\bar
   \chi_iT_{ab}{}^{ij}\gamma^b\Psi_j
     +\tfrac34 \Psi_i\Slash{\Lambda}_{\mathrm{K}}\Slash{D}
     \chi^i \,, \nonumber \\[.2ex]
     \delta_\mathrm{K} \Box_\mathrm{c}\Slash{D}\Psi_i =&\,
     \Slash{\Lambda}_{\mathrm{K}}\big[\tfrac14\big(
     R(\mathcal{V})_{ab}{}^j{}_i + 2\mathrm{i} R(A)_{ab}\,\delta^j{}_i
     \big)\gamma^{ab}\Psi_j  - \tfrac32 D\,\Psi_i\big]  \nonumber \\
  &\,
  + \Slash{\Lambda}_{\mathrm{K}}\big[
   \tfrac32 B_{ij} \,\chi^j
   - \varepsilon_{ij} F^{-\,ab} \, R(Q)^j_{ab}
   -\tfrac34 \varepsilon_{ij}
   F_{ab}^- \,  \gamma^{ab} \chi^j \big]\,,
\end{align}
These results follow from \eqref{eq:K-var-cubic-quartic}, upon making
use of the relevant curvatures.

\section{Invariant higher-derivative couplings}
\label{sec:inv-higher-der-coulings}
\setcounter{equation}{0}
Using the results of the previous section one can construct a large
variety of superconformal invariants for chiral multiplets with
higher-derivative couplings. For unrestricted chiral supermultiplets
one cannot write down Lagrangians that are at most quadratic in
derivatives, so they usually play a role as composite fields that are
expressed in terms of reduced chiral multiplets, such as the vector
multiplets and the Weyl multiplet. The construction of the higher-order
Lagrangians therefore proceeds in two steps. First one constructs the
Lagrangian in terms of unrestricted chiral multiplets of the
appropriate Weyl weights, and subsequently one expresses the
unrestricted supermultiplets in terms of reduced supermultiplets. In
these expressions it is natural to introduce a variety of arbitrary
homogeneous functions.

The invariants are expressed as chiral superspace integrals, because
all possible anti-chiral fields are contained in the kinetic
multiplets that we have introduced in section
\ref{sec:kinet-chir-mult}. A simple example of this approach was
already exhibited in (\ref{eq:real-chiral-n2}). The fact that these
invariants are actually based on full superspace integrals implies
that they must vanish whenever all the chiral (or, alternatively, all
the anti-chiral) fields are put equal to a constant. In the chiral
formulation of the integral, this phenomenon is reflected in the fact
that the kinetic multiplet of a {\it constant} anti-chiral multiplet
vanishes. This result can easily be deduced from
(\ref{eq:T-components}). Invariants can be substantially more
complicated than (\ref{eq:real-chiral-n2}). The integrand does not
have to be linear in a kinetic multiplet, and can depend on a function
of kinetic multiplets. One can also consider `nested' situations,
where a kinetic multiplet is constructed starting from an expression
of superfields among which there are other kinetic multiplets, thus
leading to even higher multiple derivatives.

The above approach is a constructive one and in general it will be
hard to classify all these invariant couplings, say, in terms of a
limited number of functions, as is often possible for supersymmetric
theories. For definiteness, we henceforth restrict attention to
invariants proportional to a single kinetic multiplet, as given in
(\ref{eq:real-chiral-n2}). In that case, expressing the composite
chiral multiplets in terms of vector multiplets, one obtains the
supergravity-coupled invariants corresponding to the actions derived
in \cite{Henningson:1995eh,deWit:1996kc} in the abelian limit, which
contain $F^4$-couplings. By including the Weyl multiplet, one also
obtains $R^2F^2$- and $R^4$-couplings. The $R^2F^2$-couplings will in
principle overlap with part of a subclass of invariants discussed
in \cite{Morales:1996bp, Antoniadis:2010iq} in connection with certain
deformations of the topological string partition function.  These
couplings are encoded in terms of a single function of holomorphic and
anti-holomorphic fields. In a rigid supersymmetry background these
actions exhibit K\"ahler geometry with this function playing the role
of a K\"ahler potential, just as happens in $N=1$ supersymmetric
actions for non-linear sigma models. As we will demonstrate below,
this feature survives in the presence of the superconformal
background. Other examples of higher-derivative couplings based on
more than a single kinetic multiplet will be discussed in
section~\ref{sec:conclusions}.

Hence we start by writing down the bosonic terms of the Lagrangian
(\ref{eq:real-chiral-n2}). It is convenient to first note the
following relation,
\begin{align}
  \label{eq:total-der-kinetic-C}
  C\vert_{\mathbb{T}(\bar\Phi)}=&\,
  \tfrac1{16} (T_{abij}\varepsilon^{ij})^2 \bar C +
  4\,\big(\mathcal{D}^\mu\mathcal{D}_\mu\big)^2\bar A \nonumber \\[.1ex]
    &\,
  - 8\,\mathcal{D}^\mu\big[\big (R_\mu{}^a(\omega,e) -\tfrac13
  R(\omega,e)\,e_\mu{}^a  - D\, e_\mu{}^a + \mathrm{i}
  R(A)_\mu{}^a\big) \mathcal{D}_a\bar
  A\big]  \nonumber\\[.1ex]
  &\,
  +\mathcal{D}_\mu\big[\varepsilon^{ij} \mathcal{D}^\mu T_{bc ij}\,F^{+bc} +4
  \,\varepsilon^{ij} T^{\mu b}{}_{ij} \,\mathcal{D}^cF^{+}_{cb} - 2\,
     T_{bc}{}^{ij}\, T^{\mu c}{}_{ij} \,\mathcal{D}^b\bar A \big]
     \nonumber\\[.1ex]
     &\,   +\cdots\,,
\end{align}
where we suppressed all fermionic contributions. In deriving this
result we made use of (\ref{eq:f-bos-uncon}). Subsequently we derive
the bosonic part of the Lagrangian corresponding to
(\ref{eq:real-chiral-n2}), making use of the density formula
\eqref{eq:chiral-density} and of the product rule
\eqref{eq:chiral-mult-prod},
\begin{align}
  \label{eq:quadratic-chiral-Lagr}
  e^{-1}\mathcal{L} =&\,
  4\,\mathcal{D}^2 A\,\mathcal{D}^2\bar A
  + 8\,\mathcal{D}^\mu A\, \big[R_\mu{}^a(\omega,e) -\tfrac13
  R(\omega,e)\,e_\mu{}^a \big]\mathcal{D}_a\bar A + C\,\bar C
  \nonumber \\[.1ex]
  &\,
   - \mathcal{D}^\mu B_{ij} \,\mathcal{D}_\mu B^{ij} + (\tfrac16
   R(\omega,e) +2\,D) \,
   B_{ij} B^{ij} \nonumber\\[.1ex]
   &\,
   - \big[\varepsilon^{ik}\,B_{ij} \,F^{+\mu\nu} \,
   R(\mathcal{V})_{\mu\nu}{}^{j}{}_{k} +\varepsilon_{ik}\,B^{ij}
   \,F^{-\mu\nu} R(\mathcal{V})_{\mu\nu j}{}^k \big] \nonumber\\[.1ex]
  &\,
  -8\, D\, \mathcal{D}^\mu A\, \mathcal{D}_\mu\bar A + \big(8\, \mathrm{i}
  R(A)_{\mu\nu} +2\, T_\mu{}^{cij}\, T_{\nu cij}\big) \mathcal{D}^\mu
  A \,\mathcal{D}^\nu\bar A  \nonumber\\[.1ex]
  &\,
  -\big[\varepsilon^{ij} \mathcal{D}^\mu T_{bc ij}\mathcal{D}_\mu
  A\,F^{+bc}+ \varepsilon_{ij} \mathcal{D}^\mu T_{bc}{}^{ij} \mathcal{D}_\mu
  \bar A\,F^{-bc}\big] \nonumber\\[.1ex]
  &\,
  -4\big[\varepsilon^{ij} T^{\mu b}{}_{ij}\,\mathcal{D}_\mu A
  \,\mathcal{D}^cF^{+}_{cb} + \varepsilon_{ij} T^{\mu
    bij}\,\mathcal{D}_\mu \bar A \,\mathcal{D}^cF^{-}_{cb}\big]
     \nonumber\\[.1ex]
    &\, + 8\, \mathcal{D}_a F^{-ab}\, \mathcal{D}^c F^+{}_{cb}  + 4\,
    F^{-ac}\, F^+{}_{bc}\, R(\omega,e)_a{}^b
     +\tfrac1{4} T_{ab}{}^{ij} \,T_{cdij} F^{-ab} F^{+cd}  \,.
\end{align}
Note that we suppressed the prime on the second chiral multiplet
indicated in \eqref{eq:real-chiral-n2}. In general, however, we will
not always identify the two multiplets, so that the complex conjugated
components in the above formula do not have to correspond to the same
supermultiplet. However, upon making this identification, the above
Lagrangian is manifestly real, which provides an additional check on
the correctness of our result. The reason is that the corresponding
lowest-order Lagrangian \eqref{eq:real-chiral-n2} is also real in that
case (up to total derivatives that we have also suppressed in deriving
the above result). Note also that the Lagrangian
\eqref{eq:quadratic-chiral-Lagr} vanishes whenever either one of the
multiplets is equal to a constant, thus confirming the analysis
presented at the beginning of this section.

We will now use the above results to write down the extension to local
supersymmetry of the class of vector multiplet Lagrangians constructed
in \cite{Henningson:1995eh,deWit:1996kc}. Just as above we concentrate
on the purely bosonic terms. The extension follows by writing the
$w=0$ chiral multiplets $\Phi$ and $\Phi^\prime$ as composite
multiplets expressed in terms of vector multiplets. In
(\ref{eq:real-chiral-n2}), and correspondingly in
\eqref{eq:quadratic-chiral-Lagr}, one thus performs the following
substitutions,
\begin{equation}
  \label{eq:substitPhi}
  \Phi\to f(\Phi^I)\,,\qquad \bar\Phi^\prime \to \bar g(\bar
  \Phi^{I})\,,
\end{equation}
where $\Phi^I$ denote the (reduced) chiral multiplets associated with
vector multiplets. Upon expanding $\Phi$ and $\bar\Phi^\prime$ in
terms of the vector supermultiplets, making use of the material
presented in appendices \ref{App:chiral-multiplets} and
\ref{App:reduced-multiplets}, one obtains powers of the vector
multiplet components multiplied by derivatives of $f(X)$ and $\bar
g(\bar X)$, where the $X^I$ denote the complex scalars of the vector
multiplets. Homogeneity implies that $X^I\,f_I(X)= 0= \bar X^I\, \bar
g_{\bar I}(\bar X)$, where $f_I$ and $\bar g_{\bar I}$ denote the
first derivatives of the two functions with respect to $X^I$ and $\bar
X^I$, respectively. Here we recall that the expression
\eqref{eq:quadratic-chiral-Lagr} vanishes whenever $f(X)$ or $\bar
g(\bar X)$ are constant. As noted previously, the origin of this
phenomenon can be traced back to the fact that the full superspace
integral of a chiral or an anti-chiral field vanishes (up to total
derivatives). Therefore the Lagrangian will depend exclusively on
mixed holomorphic/anti-holomorphic derivatives of the product function
$f(X)\,\bar g(\bar X)$. By summing over an arbitrary set of pairs of
functions $f^{(n)}(X)\, \bar g^{(n)}(\bar X)$, we can further extend
this function to a general function $\mathcal{H}(X,\bar X)$ that is
separately homogeneous of zeroth degree in $X$ and $\bar X$. Because
$\mathcal{H}(X,\bar X)$ is only defined up to a purely holomorphic or
anti-holomorphic function, it is thus subject to K\"ahler
transformations,
\begin{equation}
  \label{eq:kahler}
  \mathcal{H}(X,\bar X)\to \mathcal{H}(X,\bar X) +
  \Lambda(X)+\bar\Lambda(\bar X)\,.
\end{equation}
Hence $\mathcal{H}(X,\bar X)$ can be regarded as a K\"ahler potential,
which may be taken real (so that $\bar\Lambda(\bar
X)=[\Lambda(X)]^\ast$).

Carrying out the various substitutions leads directly to the following
bosonic contribution to the supersymmetric Lagrangian (for
convenience, we assume $\mathcal{H}$ to be real, unless stated
otherwise),
\begin{align}
  \label{eq:real-susp-action}
  e^{-1}\mathcal{L} =&\, \mathcal{H}_{IJ\bar K \bar L}\Big[\tfrac14
    \big( F_{ab}^-{}^I\, F^{-ab\,J}
                -\tfrac12 Y_{ij}{}^I\, Y^{ijJ} \big)
                \big( F_{ab}^+{}^K \, F^{+ab\,L} -\tfrac12 Y^{ijK}\,
                  Y_{ij}{}^L  \big)
              \nonumber\\
              & \qquad\quad +4\,\mathcal{D}_a X^I\, \mathcal{D}_b \bar X^K
                \big(\mathcal{D}^a X^J \,\mathcal{D}^b \bar X^L
                  + 2\, F^{-\,ac\,J}\,F^{+\,b}{}_c{}^L -
                  \tfrac14 \eta^{ab}\, Y^J_{ij}\,Y^{L\,ij}\big)
              \Big]\nonumber\\[.5ex]
   +&\,\Big\{ \mathcal{H}_{IJ\bar K}\Big[4\,\mathcal{D}_a X^I\,
     \mathcal{D}^a X^J\, \mathcal{D}^2\bar X^K
      - \big(F^{-ab\,I}\, F_{ab}^{-\,J} -\tfrac12 Y^I_{ij}\, Y^{Jij})
      \big( \Box_\mathrm{c} X^K + \tfrac18 F^{-\,K}_{ab}\, T^{ab ij}
          \varepsilon_{ij}\big)  \nonumber\\
  & \qquad\quad +8 \,\mathcal{D}^a X^I F^{-\,J}_{ab}
  \big( \mathcal{D}_cF^{+\,cb\,K}- \tfrac12 \mathcal{D}_c\bar X^K
              T^{ij\,cb} \varepsilon_{ij}\big) - \mathcal{D}_a
            X^I\, Y^J_{ij}\,\mathcal{D}^aY^{K\,ij}\Big]
            +\mathrm{h.c.}\Big\}   \nonumber\\[.5ex]
     +&\mathcal{H}_{I\bar J}\Big[ 4\big( \Box_\mathrm{c} \bar X^I + \tfrac18
         F_{ab}^{+\,I}\, T^{ab}{}_{ij} \varepsilon^{ij}\big)
     \big( \Box_\mathrm{c}  X^J + \tfrac18 F_{ab}^{-\,J}\, T^{abij}
       \varepsilon_{ij}\big) + 4\,\mathcal{D}^2 X^I \,\mathcal{D}^2
       \bar X^J \nonumber\\
       & \quad\quad +8\,\mathcal{D}_{a}F^{-\,abI\,}\,
       \mathcal{D}_cF^{+c}{}_{b}{}^J   - \mathcal{D}_a Y_{ij}{}^I\,
            \mathcal{D}^a Y^{ij\,J}
            +\tfrac1{4} T_{ab}{}^{ij} \,T_{cdij}
            \,F^{-ab\,I}F^{+cd\,J}
     \nonumber\\
     &\quad\quad
     +\big(\tfrac16 R(\omega,e) +2\,D\big) Y_{ij}{}^I\, Y^{ij\,J}   + 4\,
     F^{-ac\,I}\, F^{+}{}_{bc}{}^J \, R(\omega,e)_a{}^b  \nonumber\\
     &\quad\quad + 8\big(R^{\mu\nu}(\omega,e)-\tfrac13 g^{\mu\nu}
     R(\omega,e) +\tfrac1{4} T^\mu{}_{b}{}^{ij}\, T^{\nu b}{}_{ij}
     +\mathrm{i} R(A)^{\mu\nu} - g^{\mu\nu} D\big) \mathcal{D}_\mu X^I
     \,\mathcal{D}_\nu \bar X^J  \nonumber\\
     &\quad\quad
     - \big[\mathcal{D}_c \bar X^J \big(\mathcal{D}^c
     T_{ab}{}^{ij}\,F^{-\,I\,ab} +4
       \,T^{ij\,cb} \,\mathcal{D}^aF^{-\,I}_{ab} \big)\varepsilon_{ij}
       +[\mathrm{h.c.}; I\leftrightarrow J]  \big]\nonumber\\
     &\quad\quad -\big[\varepsilon^{ik}\, Y_{ij}{}^I\, F^{+ab\,J}\,
      R(\mathcal{V})_{ab}{}^j{}_k +[\mathrm{h.c.}; I\leftrightarrow J]
      \big]  \Big] \,,
\end{align}
where (we suppress fermionic contributions),
\begin{align}
  \label{eq:def}
  F_{ab}^-{}^I =&\,  \big(\delta_{ab}{}^{cd} -\tfrac12
    \varepsilon_{ab}{}^{cd}\big) e_c{}^\mu e_d{}^\nu \,\partial_{[\mu}
    W_{\nu]}{}^I
  -\tfrac14\, \bar{X}^I\, T_{ab}{}^{ij}\varepsilon_{ij} \,,
  \nonumber\\
  \Box_\mathrm{c} X^I=&\, \mathcal{D}^2 X^I + \big(\tfrac16
  R(\omega,e) +D\big) \,X^I \,.
\end{align}
In view of the K\"ahler equivalence transformations \eqref{eq:kahler},
the mixed derivative $\mathcal{H}_{I\bar J}$ can be identified as a
K\"ahler metric. Hence we have the following results for the metric,
connection, and the curvature of the corresponding K\"ahler space,
\begin{align}
  \label{eq:K-geom}
  g_{I\bar J}=&\,\mathcal{H}_{I\bar J} \,, \nonumber\\
  \Gamma^I{}_{JK} =&\,g^{I\bar L} \,\mathcal{H}_{JK\bar L}\,,\nonumber\\
  R_{I\bar J K\bar L}=&\, \mathcal{H}_{IK\bar J \bar L} - g_{M\bar N}\,
  \Gamma^M{}_{IK} \,\Gamma^{\bar N}{}_{\bar J\bar L} \,.
\end{align}
The Lagrangian \eqref{eq:real-susp-action} can then be written in a
K\"ahler covariant form,
\begin{align}
  \label{eq:kahler-susp-action}
  e^{-1}\mathcal{L} =&\, R_{I\bar K J\bar L}\Big[\tfrac14
    \big( F_{ab}^-{}^I\, F^{-ab\,J}
                -\tfrac12 Y_{ij}{}^I\, Y^{ijJ} \big)
                \big( F_{ab}^+{}^K \, F^{+ab\,L} -\tfrac12 Y^{ijK}\,
                  Y_{ij}{}^L  \big)
              \nonumber\\
              & \qquad\quad +4\,\mathcal{D}_a X^I\, \mathcal{D}_b \bar X^K
                \big(\mathcal{D}^a X^J \,\mathcal{D}^b \bar X^L
                  + 2\, F^{-\,ac\,J}\,F^{+\,b}{}_c{}^L -
                  \tfrac14 \eta^{ab}\, Y^J_{ij}\,Y^{L\,ij}\big)
                  \Big]\nonumber\\[.5ex]
     &+g_{I\bar J}\Big[ 4\big( \Box_\mathrm{c} \bar X^I +
     \tfrac18 F_{ab}^{+\,I}\, T^{ab}{}_{ij}
     \varepsilon^{ij}-\tfrac14\Gamma^I{}_{KL}( F_{ab}^-{}^K\,
     F^{-ab\,L} -\tfrac12 Y^{ijK}\, Y{}_{ij}{}^L) \big)  \nonumber\\
     &\qquad\qquad\times
     \big( \Box_\mathrm{c}  X^J + \tfrac18 F_{ab}^{-J}\, T^{abij}
     \varepsilon_{ij} -\tfrac14\Gamma^{\bar J}{}_{\bar K\bar L}( F_{ab}^+{}^K\,
     F^{+ab\,L} -\tfrac12 Y^{ijK}\, Y{}_{ij}{}^L)\big) \nonumber \\
       &\quad\quad + 4\,\big(\mathcal{D}^2 X^I+\Gamma^{I}{}_{KL}\,
       \mathcal{D}_b X^K \,\mathcal{D}^b X^L \big)
       \,\big(\mathcal{D}^2\bar X^J +\Gamma^{\bar J}{}_{\bar K\bar L}  \,
       \mathcal{D}_b \bar X^K \,\mathcal{D}^b \bar X^L \big)
       \nonumber\\
       & \quad\quad +8\,\big(\mathcal{D}_{a}F^{-\,abI}
       +\Gamma^{I}{}_{KL}  \, \mathcal{D}_a X^K F^{-\,abL} \,\big)\,
       \big(\mathcal{D}_cF^{+c}{}_{b}{}^J+\Gamma^{\bar
       J}{}_{\bar K\bar L}  \, \mathcal{D}_c \bar X^K \,F^{+c}{}_{b}{}^L \big)
     \nonumber\\
     &\quad\quad - (\mathcal{D}_a Y_{ij}{}^I+\Gamma^{I}{}_{KL}  \,
     \mathcal{D}_b  X^K \,Y_{ij}{}^L \big)\,
            \big(\mathcal{D}^a Y^{J\,ij}+\Gamma^{\bar J}{}_{\bar K\bar
            L}  \, \mathcal{D}_b \bar X^K \,Y^{ijL}\big)
     \nonumber \\
     &\quad\quad
     +\tfrac1{4} T_{ab}{}^{ij} \,T_{cdij}
            \,F^{-ab\,I}F^{+cd\,J}
     \nonumber\\
     &\quad\quad
     +\big(\tfrac16 R(\omega,e) +2\,D\big) Y_{ij}{}^I\, Y^{ij\,J}   + 4\,
     F^{-ac\,I}\, F^{+}{}_{bc}{}^J \, R(\omega,e)_a{}^b  \nonumber\\
     &\quad\quad + 8\big(R^{\mu\nu}(\omega,e)-\tfrac13 g^{\mu\nu}
     R(\omega,e) +\tfrac1{4} T^\mu{}_{b}{}^{ij}\, T^{\nu b}{}_{ij}
     +\mathrm{i} R(A)^{\mu\nu} - g^{\mu\nu} D\big) \mathcal{D}_\mu X^I
     \,\mathcal{D}_\nu \bar X^J  \nonumber\\
     &\quad\quad
     - \big[\mathcal{D}_c \bar X^J \big(\mathcal{D}^c
     T_{ab}{}^{ij}\,F^{-\,I\,ab} +4 \,T^{ij\,cb}
       \,(\mathcal{D}^aF^{-\,I}_{ab}+\Gamma^I{}_{KL}\mathcal{D}^a X^K
       F^-_{ab}{}^L)  \big)\varepsilon_{ij}
       +[\mathrm{h.c.}; I\leftrightarrow J] \big]\nonumber\\
       &
      \quad\quad -\big[\varepsilon^{ik}\, Y_{ij}{}^I\, F^{+ab\,J}\,
      R(\mathcal{V})_{ab}{}^j{}_k +[\mathrm{h.c.}; I\leftrightarrow
      J]\big]  \Big] \,.
\end{align}
The covariantizations in the various combinations can be understood
systematically by rewriting the chiral multiplet components of the
vector multiplets such that they are covariant with respect to the
complex reparametrizations of the K\"ahler space (in the limit where
the fermions are suppressed). An easy way to appreciate these
covariantizations is by reorganizing the expansion of a composite
chiral multiplet into vector multiplets according to
\eqref{eq:chiral-mult-exp} by replacing the ordinary derivatives of the
function $\mathcal{G}$ by covariant derivatives.

The Lagrangians \eqref{eq:real-susp-action} and/or
\eqref{eq:kahler-susp-action} can also be used in the context of rigidly
supersymmetric theories upon suppressing all the superconformal
fields. The resulting Lagrangian is then superconformally invariant in
flat Minkowski space. This invariance can be further reduced to ordinary
Poincar\'e supersymmetry by replacing one of the vector multiplets by
a constant.

As an extension of the previous results we return to
\eqref{eq:quadratic-chiral-Lagr}, and consider composite chiral
multiplets that depend on both vector multiplets and on the Weyl
multiplet. Hence we replace \eqref{eq:substitPhi} by
\begin{equation}
  \label{eq:substiPhiWeyl}
  \Phi\to f(\Phi^I, W^2) \,,\qquad  \bar\Phi^\prime\to \bar
  g(\bar\Phi^{I},  \bar W^2) \,,
\end{equation}
where $W^2$ refers to the square of the Weyl multiplet. The components
of this reduced chiral multiplet are given in
\eqref{eq:W-squared}. Upon expanding these functions and substituting
the results into \eqref{eq:quadratic-chiral-Lagr}, one obtains a
Lagrangian that contains $R^4$-, $R^2 F^2$- and $F^4$-terms. All terms
are proportional to mixed holomorphic/anti-holomorphic derivatives of
a function $\mathcal{H}(X,T^2,\bar X,\bar T^2)$, where
$T^2=(T_{ab}{}^{ij}\varepsilon_{ij})^2$ and $\bar
T^2=(T_{abij}\varepsilon^{ij})^2$, and where $\mathcal{H}$ is
constructed from pairs of products of functions $f(X,T^2)$ and $\bar
g(\bar X,\bar T^2)$. The fact that the composite multiplets have $w=0$
implies a modified homogeneity property,
\begin{equation}
  \label{eq:mod-hom}
  X^I\mathcal{H}_I(X,T^2,\bar X,\bar T^2) + 2\, T^2
  \mathcal{H}_{T^2}(X,T^2,\bar X,\bar T^2)=0 \,,
\end{equation}
and likewise for the anti-holomorphic derivatives.

The Lagrangian consists of the Lagrangian \eqref{eq:real-susp-action}
plus a large number of terms that involve multiple derivatives of
$\mathcal{H}$ with respect to $T^2$, $\bar T^2$, $X^I$ and $\bar
X^I$. Below we concentrate on terms proportional to multiple
derivatives of $\mathcal{H}$ with respect to only $T^2$ and $\bar
T^2$. Among others those contain contributions of fourth order in
$\mathcal{R}(M)$, whose leading contribution is equal to the Weyl
tensor,
\begin{align}
  \label{eq:add-R4}
  &(64)^{-2}\,e^{-1}\mathcal{L} =  \nonumber\\
  &\quad
  4\,\mathcal{H}_{T^2T^2\bar T^2\bar T^2} \; T^{abij} \varepsilon_{ij}\,
  T^{cdkl}\varepsilon_{kl}\, T^{ef}{}_{mn}\varepsilon^{mn}\,
  T^{gh}{}_{pq}\varepsilon^{pq}\nonumber\\
  &\qquad\times
  \big[{\cal R}(M)_{aba'b'} {\cal R}(M)_{cd}{}^{\!a'b'}
            +\tfrac12 R({\cal V})_{ab}{}^i{}_{j}\, R({\cal
              V})_{cd}{}^j{}_{i} \big] \nonumber\\
  &\qquad\times
  \big[{\cal R}(M)_{efe'f'} {\cal R}(M)_{gh}{}^{\!e'f'}
            +\tfrac12 R({\cal V})_{ef}{}^i{}_{j}\, R({\cal
              V})_{gh}{}^j{}_{i} \big] \nonumber\\[.5ex]
     &\quad+ 2\,\Big\{\mathcal{H}_{T^2T^2\bar T^2} \; T^{abij}
     \varepsilon_{ij}\, T^{cdkl}\varepsilon_{kl}\nonumber\\
     &\qquad\times
     \big[{\cal R}(M)_{aba'b'} {\cal R}(M)_{cd}{}^{\!a'b'}
            +\tfrac12 R({\cal V})_{ab}{}^i{}_{j}\, R({\cal
              V})_{cd}{}^j{}_{i} \big] \nonumber\\
     &\qquad\times
     \big[{\cal R}(M)^+_{efgh} {\cal R}(M)^{+ efgh}
            +\tfrac12 R({\cal V})^+_{ef}{}^i{}_{j}\, R({\cal
              V})^{+efj}{}_{i} -\tfrac12 T^{ef}{}_{mn}D_eD^hT_{hf}{}^{mn}\big]
   + [\mathrm{h.c.}] \,\Big\} \nonumber\\[.5ex]
    &\quad +\mathcal{H}_{T^2\bar T^2} \Big\{\big\vert{\cal R}(M)^+_{abcd}
    {\cal R}(M)^{+abcd}
            +\tfrac12 R({\cal V})^+_{ab}{}^i{}_{j}\, R({\cal
              V})^{+abj}{}_{i} -\tfrac12
            T^{ab}{}_{mn}D_aD^eT_{eb}{}^{mn}\big\vert^2
                    + \cdots \Big\}\,.    \nonumber\\
    &\qquad\qquad {~}
\end{align}
Besides the terms quartic in $\mathcal{R}(M)$ we have retained some of
the terms that come with them as part of the basic building blocks
that emerge in the calculation (similar blocks appear
in~\eqref{eq:real-susp-action}). Besides giving a little more
information in this way, this has the advantage that the origin of the
various term will be easier to track down.

In addition to the above terms there are mixed terms which lead to
explicit contributions from the vector multiplets (i.e. beyond the $X$
and $\bar X$ dependence in the function $\mathcal{H}$). Those include,
for instance, terms proportional to $[\mathcal{R}(M)]^2$ times the
product of two vector multiplet field strengths, $F_{\mu\nu}{}^I$. We
will not exhibit those terms here (they can in principle be deduced
from \eqref{eq:quadratic-chiral-Lagr} along the same lines as for the
previous contributions). Some of these terms will be shown in the
equation below.

A special case, which is worth mentioning in view of the work of
\cite{Morales:1996bp, Antoniadis:2010iq}, corresponds to functions
$\mathcal{H}(X,T^2,\bar X)$ that do not depend on $\bar T^2$. Hence
the function $\mathcal{H}$ is not real. Again we do not present all
the terms, but we give all the terms that contain $\mathcal{R}(M)$
(with some completions), with the exception of terms proportional to
derivatives of $X^I$ and $T_{ab}{}^{ij}$ or their complex conjugates,
\begin{align}
  \label{eq:holo-T2}
  &(64)^{-1}e^{-1}\mathcal{L}= \nonumber\\
       &\quad
       \mathcal{H}_{T^2 T^2\bar K \bar L}\Big\{
       T^{abij} \varepsilon_{ij}\,
       T^{cdkl}\varepsilon_{kl}\,
       \big[{\cal R}(M)_{aba'b'} {\cal R}(M)_{cd}{}^{\!a'b'}
            +\tfrac12 R({\cal V})_{ab}{}^i{}_{j}\, R({\cal
              V})_{cd}{}^j{}_{i} \big] \nonumber\\
            &\quad\quad\qquad\qquad \times \big[F^{+}_{ef}{}^K
            F^{+efL} -\tfrac12 Y^{mn}{}^K Y_{mn}{}^L \big]
            + \cdots \Big\}  \nonumber\\[.5ex]
       & \quad -4\, \mathcal{H}_{T^2 T^2\bar K}\Big\{
        T^{abij} \varepsilon_{ij}\,
       T^{cdkl}\varepsilon_{kl}\,
       \big[{\cal R}(M)_{aba'b'} {\cal R}(M)_{cd}{}^{\!a'b'}
            +\tfrac12 R({\cal V})_{ab}{}^i{}_{j}\, R({\cal
              V})_{cd}{}^j{}_{i} \big] \nonumber\\
       &\quad\qquad\qquad\qquad \times
              \big[ \Box_\mathrm{c} X^K + \tfrac18 F_{ef}{}^K
               T^{efij} \varepsilon_{ij}\big] +\cdots \Big\}
               \nonumber\\[.5ex]
       & \quad
       + \tfrac12\mathcal{H}_{T^2 I \bar K}\Big\{
          T^{cdlm} \varepsilon_{lm} \big[F^{-}_{ab}{}^I
         \mathcal{R}(M)_{cd}{}^{ab}-\tfrac12 Y^{ijI}  \varepsilon_{ki}\,
       R(\mathcal{V})_{cd}{}^k{}_j\big] \nonumber\\
         &\quad\qquad\qquad\qquad \times
         \big[ \Box_\mathrm{c} X^K + \tfrac18 F_{ef}{}^K T^{efij}
         \varepsilon_{ij}\big] +\cdots \Big\}  \nonumber\\[.5ex]
       & \quad
       -\tfrac18\mathcal{H}_{T^2 I \bar K \bar L}\Big\{
             T^{cdlm} \varepsilon_{lm} \big[F_{ab}{}^I
         \mathcal{R}(M)_{cd}{}^{ab}-\tfrac12 Y^{ijI}  \varepsilon_{ki}\,
       R(\mathcal{V})_{cd}{}^k{}_j\big] \nonumber\\
       &\quad\qquad\qquad\qquad \times
       \big[  F^{+}_{ab}{}^I F^{+abJ} -\tfrac12 Y^{ij}{}^K
       Y_{ij}{}^L \big] +\cdots \Big\} \nonumber\\[.5ex]
      &\quad
      + \tfrac12\mathcal{H}_{T^2\bar K \bar L}\Big\{
      \big[\mathcal{R}(M)^-_{cdef}\mathcal{R}(M)^{-cdef}
            +\tfrac12 R({\cal V})^-_{cd}{}^i{}_{j}\, R({\cal
              V})^{-cdj}{}_{i} -\tfrac12 T^{cd mn}D_cD^eT_{ed mn}\big]
            \nonumber\\
       &\quad\qquad\qquad\qquad
       \times \big[F^{+}_{ab}{}^I F^{+abJ} -\tfrac12 Y^{ijK}
       Y_{ij}{}^L \big] +\cdots \Big\}
                \nonumber\\[.5ex]
       &\quad
       -2\, \mathcal{H}_{T^2\bar K } \Big\{
       \big[ \mathcal{R}(M)^-_{abcd}\mathcal{R}(M)^{-abcd}
            +\tfrac12 R({\cal V})^-_{ab}{}^i{}_{j}\, R({\cal
              V})^{-abj}{}_{i} -\tfrac12 T^{ab mn}D_aD^cT_{cb mn}\big]
            \nonumber\\
        &\quad\qquad\qquad\qquad  \times
              \big[\Box_\mathrm{c} X^K + \tfrac18 F_{ef}{}^K
               T^{ef ij} \varepsilon_{ij}\big] \nonumber\\
        &\quad\qquad\qquad
        + T_{cd}{}^{ij}\varepsilon_{ij} \,\mathcal{R}(M)^{cdab}\big[
        \tfrac1{32} T_{ab}{}^{kl} T_{efkl} \, F^{efK}
        +\tfrac12 F_{eb}^+{}^K \, R(\omega,e)_a{}^e
          -\tfrac1{8}\varepsilon_{km}\,  Y^{kl}{}^K
          R(\mathcal{V})_{ab}{}_{l}{}^{m}\big]
         \nonumber\\
        &\quad\qquad\qquad
        + T_{cd}{}^{ij}\varepsilon_{ij} \, \mathcal{D}_a
        \mathcal{R}(M)^{cdab}  \, \mathcal{D}^e F_{eb}^+{}^K  +\cdots
        \Big\}\,.
\end{align}

\section{A non-renormalization theorem for BPS black hole entropy}
\label{sec:non-renorm-theor-BPS}
\setcounter{equation}{0}
The results of this paper can be used in the study of black holes.
Based on any linear combination of the various $N=2$ locally
supersymmetric Lagrangians, one can evaluate the corresponding
expressions for the Wald entropy and the electric charges in terms of
the values of the fields taken at the black hole horizon. In the case
of BPS black holes, the horizon values of the fields are highly
restricted due to full supersymmetry enhancement at the horizon, and
therefore the resulting expressions for the entropy and the charges
will simplify. To explore this one must determine the possible
supersymmetric field configurations, preferably in an off-shell
formulation so that the results do not depend on the specific
Lagrangian. This has already been done in \cite{LopesCardoso:2000qm},
which provided a generalization of the attractor equations found in
\cite{Ferrara:1995ih, Strominger:1996kf, Ferrara:1996dd}. So far,
generic chiral supermultiplets were not considered, but it is
convenient to do so as well. As it will turn out, it suffices
to restrict oneself to chiral multiplets of Weyl weight $w=0$, for
which results are rather straightforward to obtain.

The first relevant observation is that a constant chiral superfield
(i.e. a supermultiplet with constant $A$ and all other components
vanishing) is only supersymmetric provided it has $w=0$. In fact there
exist no other supersymmetric values of the chiral superfield. All
this can be derived directly from the transformation rules
(\ref{eq:conformal-chiral}). The second observation is that the
kinetic multiplet constructed from a $w=0$ anti-chiral multiplet,
vanishes when the latter multiplet is equal to a constant. This
follows by inspection of~(\ref{eq:T-components}). These two
observations prove immediately that any invariant proportional to a
kinetic multiplet, must vanish for supersymmetric field
configurations. This fact can immediately be verified from
(\ref{eq:quadratic-chiral-Lagr}), because when the fields $A$ and
$\bar A^\prime$ are constant and all other chiral multiplet component
fields are vanishing, the expression (\ref{eq:quadratic-chiral-Lagr})
indeed vanishes.

The above result is interesting in its own right, but we are also
interested in the first-order variation of the action induced by a
change of some of the fields, evaluated for a supersymmetric
background. Given the fact that all the invariants discussed in this
paper will contain at least one kinetic multiplet, we thus consider
\begin{equation}
  \label{eq:variation}
  \delta\mathcal{L} \propto \int \;\mathrm{d}^4\theta \,\Big[
  \delta\Phi\,\mathbb{T}(\bar\Phi^\prime) +
  \Phi\,\delta\mathbb{T}(\bar\Phi^\prime) \Big] \,,
\end{equation}
where $\Phi$ and $\Phi^\prime$ are composite chiral fields, which are
themselves expressed in various chiral fields, including possible
kinetic multiplets. They are not necessarily uniquely defined, and it
is also possible to consider linear combinations of such terms. Since we
will be evaluating the variation at supersymmetric values of the
fields, the first term in (\ref{eq:variation}) vanishes, because the
kinetic multiplet vanishes, whereas the second term can be evaluated
for constant $\Phi$.

However, rather than continuing in this way, we may simply return to
(\ref{eq:quadratic-chiral-Lagr}) and consider its variation. Observe
that each term is proportional to a product of one component of $\Phi$
and another one of $\bar \Phi^\prime$ (we remind the reader that in
(\ref{eq:quadratic-chiral-Lagr}) we suppressed the prime for
notational clarity). All these components will be equal to zero in a
supersymmetric background, with the exception of $A$ and $\bar
A^\prime$, which will take constant values. However, only space-time
derivatives of $A$ and $\bar A^\prime$ appear, and those will vanish as
well. In other words, (\ref{eq:quadratic-chiral-Lagr}) is always
quadratic in quantities that are vanishing in the supersymmetry
limit. Hence any first-order variation of any Lagrangian of this type
must necessarily vanish in a supersymmetric background!

The above result suffices to derive a non-renormalization theorem for
electric charges and the Wald entropy
\cite{Wald:1993nt,Jacobson:1993vj,Iyer:1994ys} for BPS black
holes. The reason is that these quantities are always expressed in
terms of first-order derivatives of the Lagrangian with respect to
certain fields, such as the abelian field strengths or the Riemann
tensor, or possible derivatives thereof. This concludes the proof of
the non-renormalization theorem.

As we already mentioned in section \ref{sec:introduction}, the
existence of this non-renormalization theorem is a welcome result. So
far good agreement has been established for BPS black hole entropy
evaluated on the basis of supergravity and of microstate counting,
suggesting that other invariants in supergravity should contribute
only marginally, or perhaps not at all, at the subleading level. The
result of this section lends support to this idea. Nevertheless the
possible existence of alternative supersymmetric invariants that do
not belong to the class of invariants discussed in this paper, cannot
be excluded at this stage.

\section{Concluding remarks}
\label{sec:conclusions}
\setcounter{equation}{0}
In this paper we studied a large class of $N=2$ superconformal
invariants involving higher-derivative couplings, based on full
superspace integrals. For a special subclass we have presented
explicit results for some of the bosonic terms. This is the subclass
that contains only a single kinetic multiplet.

As indicated already, there are further options. The most obvious one
is to include more kinetic multiplets, based on various composite
chiral and anti-chiral multiplets with suitable Weyl weights,
\begin{equation}
  \label{eq:polynomial-kin}
  \int \;\mathrm{d}^4\theta \;\Phi_0\,\mathbb{T}(\bar\Phi_1)\cdots
  \mathbb{T}(\bar\Phi_n) \,,
\end{equation}
where $\bar\Phi_1, \dots \bar\Phi_n$ are anti-chiral superfields of
zero weight and $\Phi_0$ is a chiral superfield of weight $w=-2(n-1)$.
This leads to actions that contain four space-time
derivatives. However, when treating the chiral multiplets as
composites of reduced chiral multiplets, one obtains invariants with
terms of $2(1+n)$ powers of field strengths and/or explicit
derivatives, i.e., $R^{2m} F^{2p}\mathcal{D}^{2(n+1-m-p)}$. The case of
$n=1$ has been dealt with in considerable detail in section
\ref{sec:inv-higher-der-coulings}. The expression of the composite
chiral multiplets in terms of the reduced ones allows again for the
presence of functions $\mathcal{H}^{(n)}$ which are subject to
a generalized version of the K\"ahler transformations noted in section
\ref{sec:inv-higher-der-coulings}.

As alluded to before, one can also consider nested situations where
the kinetic multiplet is constructed from a combination of
(anti)chiral fields that include again other kinetic multiplets. In
this way one constructs multiplets with multiple derivatives of
arbitrary power. We are then led to introduce quantities of the type,
\begin{equation}
  \label{eq:nested-kinetic}
  \mathbb{T}^{(2)}= \mathbb{T}(\bar\Phi_2 \, \mathbb{T}(\Phi_1)) \,, \quad
  \mathbb{T}^{(3)}=\mathbb{T}(\bar\Phi_3 \, \mathbb{T}(\Phi_2 \,
  \mathbb{T}(\bar\Phi_1))) \,, \;\dots,\;
  \mathbb{T}^{(n)}= \mathbb{T}(\bar\Phi_n \,\mathbb{T}^{(n-1)}) \,,
\end{equation}
which can be part of any superspace integrand, on the same footing as
the kinetic multiplets in \eqref{eq:polynomial-kin}. Here $\Phi_1$ has
$w=0$ and $\Phi_2,\Phi_3,\cdots$ have $w=-2$. This extends the number
of invariants to all possible combinations of the form
\begin{equation}
  \label{eq:general-kin}
  \int \;\mathrm{d}^4\theta \,\Phi_0 \,\mathbb{T}^{(n_1)}\,\mathbb{T}^{(n_2)}
  \cdots \,\mathbb{T}^{(n_k)} \,,
\end{equation}
where $\Phi_0$ has $w=-2(k-1)$ and where we assume $n_k\geq1$ with
$\mathbb{T}(\bar\Phi_1)\equiv\mathbb{T}^{(1)}$. When expressing all
the chiral multiplets in terms of reduced ones, then one can show that
the maximal number of derivatives of the invariants
\eqref{eq:general-kin} is equal to $2(1+\sum_k n_k)$.

These types of invariants are not necessarily independent in the sense that
there can be linear combinations that are equal to a total
derivative. For example, at the six-derivative level, one has
\begin{equation}
  \label{eq:six-der}
  \int \;\mathrm{d}^4\theta \,\Phi_0\,\mathbb{T}(\bar\Phi_2 \,
   \mathbb{T}(\Phi_1)) \approx
  \int \;\mathrm{d}^4\bar\theta \,\bar\Phi_2\mathbb{T}(\Phi_0)\,
    \mathbb{T}(\Phi_1)  \,,
\end{equation}
up to total derivatives. Nevertheless it is clear that we are dealing
with an infinite hierarchy of higher-derivative invariants.

Of course, a relevant question is whether the invariant couplings
presented in this paper exhaust the possible higher-derivative
invariants. Most likely, this will not be the case. From the
perspective of BPS black holes the question would then remain whether
these conjectured couplings could still contribute to the entropy and
electric charges.

\section*{Acknowledgement}
We thank Warren Siegel for a useful discussion concerning the
vanishing of the full $N\!=\!2$ superspace volume. The work of
S.K. and M.v.Z. is part of the research programme of the `Stichting
voor Fundamenteel Onderzoek der Materie (FOM)', which is financially
supported by the `Nederlandse Organisatie voor Wetenschappelijk
Onderzoek (NWO)'. This work is supported in part by the ERC Advanced
Grant no. 246974, {\it ``Supersymmetry: a window to non-perturbative
  physics''}.

\begin{appendix}
%
\section{Superconformal calculus}
\label{App:SC}
\setcounter{equation}{0}
Throughout this paper we use Pauli-K\"all\'en conventions and follow
the notation used {\it e.g.} in \cite{LopesCardoso:2000qm}. Space-time
and Lorentz indices are denoted by $\mu,\nu,\ldots$, and $a,b,\ldots$,
respectively; $\mathrm{SU}(2)$-indices are denoted by $i,j,\ldots$. As
mentioned already in footnote~\ref{ftn:notation},
(anti-)sym\-metrizations are always defined with unit strength.

In this appendix we present the transformation rules of the
superconformal fields and their relation to the superconformal
algebra, as well as their covariant quantities contained in the
so-called Weyl supermultiplet. The superconformal algebra comprises
the generators of the general-coordinate, local Lorentz, dilatation,
special conformal, chiral $\mathrm{SU}(2)$ and $\mathrm{U}(1)$,
supersymmetry (Q) and special supersymmetry (S) transformations.  The
gauge fields associated with general-coordinate transformations
($e_\mu{}^a$), dilatations ($b_\mu$), chiral symmetry
($\mathcal{V}_\mu{}^i{}_j$ and $A_\mu$) and Q-supersymmetry
($\psi_\mu{}^i$) are independent fields.  The remaining gauge fields
associated with the Lorentz ($\omega_\mu{}^{ab}$), special conformal
($f_\mu{}^a$) and S-supersymmetry transformations ($\phi_\mu{}^i$) are
dependent fields.  They are composite objects, which depend on the
independent fields of the multiplet
\cite{deWit:1980tn,deWit:1984pk,deWit:1984px}. The corresponding
supercovariant curvatures and covariant fields are contained in a
tensor chiral multiplet, which comprises $24+24$ off-shell degrees of
freedom. In addition to the independent superconformal gauge fields,
it contains three other fields: a Majorana spinor doublet $\chi^i$, a
scalar $D$, and a selfdual Lorentz tensor $T_{abij}$, which is
anti-symmetric in $[ab]$ and $[ij]$. The Weyl and chiral weights have
been collected in table \ref{table:weyl}.
%
\begin{table}[t]
\begin{tabular*}{\textwidth}{@{\extracolsep{\fill}}
    |c||cccccccc|ccc||ccc| }
\hline
 & &\multicolumn{9}{c}{Weyl multiplet} & &
 \multicolumn{2}{c}{parameters} & \\[1mm]  \hline \hline
 field & $e_\mu{}^{a}$ & $\psi_\mu{}^i$ & $b_\mu$ & $A_\mu$ &
 $\mathcal{V}_\mu{}^i{}_j$ & $T_{ab}{}^{ij} $ &
 $ \chi^i $ & $D$ & $\omega_\mu^{ab}$ & $f_\mu{}^a$ & $\phi_\mu{}^i$ &
 $\epsilon^i$ & $\eta^i$
 & \\[.5mm] \hline
$w$  & $-1$ & $-\tfrac12 $ & 0 &  0 & 0 & 1 & $\tfrac{3}{2}$ & 2 & 0 &
1 & $\tfrac12 $ & $ -\tfrac12 $  & $ \tfrac12  $ & \\[.5mm] \hline
$c$  & $0$ & $-\tfrac12 $ & 0 &  0 & 0 & $-1$ & $-\tfrac{1}{2}$ & 0 &
0 & 0 & $-\tfrac12 $ & $ -\tfrac12 $  & $ -\tfrac12  $ & \\[.5mm] \hline
 $\gamma_5$   &  & + &   &    &   &   & + &  &  &  & $-$ & $ + $  & $
 -  $ & \\ \hline
\end{tabular*}
\vskip 2mm
\renewcommand{\baselinestretch}{1}
\parbox[c]{\textwidth}{\caption{\label{table:weyl}{\footnotesize
Weyl and chiral weights ($w$ and $c$) and fermion
chirality $(\gamma_5)$ of the Weyl multiplet component fields and the
supersymmetry transformation parameters.}}}
\end{table}

Under Q-supersymmetry, S-supersymmetry and special conformal
transformations the independent fields of the Weyl multiplet transform
as follows,
\begin{eqnarray}
  \label{eq:weyl-multiplet}
  \delta e_\mu{}^a & =& \bar{\epsilon}^i \, \gamma^a \psi_{ \mu i} +
  \bar{\epsilon}_i \, \gamma^a \psi_{ \mu}{}^i \, , \nonumber\\
  \delta \psi_{\mu}{}^{i} & =& 2 \,\mathcal{D}_\mu \epsilon^i - \tfrac{1}{8}
  T_{ab}{}^{ij} \gamma^{ab}\gamma_\mu \epsilon_j - \gamma_\mu \eta^i
  \, \nonumber \\
  \delta b_\mu & =& \tfrac{1}{2} \bar{\epsilon}^i \phi_{\mu i} -
  \tfrac{3}{4} \bar{\epsilon}^i \gamma_\mu \chi_i - \tfrac{1}{2}
  \bar{\eta}^i \psi_{\mu i} + \mbox{h.c.} + \Lambda^a_K e_{\mu a} \, ,
  \nonumber \\
  \delta A_{\mu} & =& \tfrac{1}{2} \mathrm{i} \bar{\epsilon}^i \phi_{\mu i} +
  \tfrac{3}{4} \mathrm{i} \bar{\epsilon}^i \gamma_\mu \, \chi_i +
  \tfrac{1}{2} \mathrm{i}
  \bar{\eta}^i \psi_{\mu i} + \mbox{h.c.} \, , \nonumber\\
  \delta \mathcal{V}_\mu{}^{i}{}_j &=& 2\, \bar{\epsilon}_j
  \phi_\mu{}^i - 3
  \bar{\epsilon}_j \gamma_\mu \, \chi^i + 2 \bar{\eta}_j \, \psi_{\mu}{}^i
  - (\mbox{h.c. ; traceless}) \, , \nonumber \\
  \delta T_{ab}{}^{ij} &=& 8 \,\bar{\epsilon}^{[i} R(Q)_{ab}{}^{j]} \,
  , \nonumber \\
  \delta \chi^i & =& - \tfrac{1}{12} \gamma^{ab} \, \Slash{D} T_{ab}{}^{ij}
  \, \epsilon_j + \tfrac{1}{6} R(\mathcal{V})_{\mu\nu}{}^i{}_j
  \gamma^{\mu\nu} \epsilon^j -
  \tfrac{1}{3} \mathrm{i} R_{\mu\nu}(A) \gamma^{\mu\nu} \epsilon^i + D
  \epsilon^i +
  \tfrac{1}{12} \gamma_{ab} T^{ab ij} \eta_j \, , \nonumber \\
  \delta D & =& \bar{\epsilon}^i \,  \Slash{D} \chi_i +
  \bar{\epsilon}_i \,\Slash{D}\chi^i \, .
\end{eqnarray}
Here $\epsilon^i$ and $\epsilon_i$ denote the spinorial parameters of
Q-supersymmetry, $\eta^i$ and $\eta_i$ those of S-supersymmetry, and
$\Lambda_K{}^a$ is the transformation parameter for special conformal
boosts.  The full superconformally covariant derivative is denoted by
$D_\mu$, while $\mathcal{D}_\mu$ denotes a covariant derivative with
respect to Lorentz, dilatation, chiral $\mathrm{U}(1)$, and
$\mathrm{SU}(2)$ transformations,
\begin{equation}
  \label{eq:D-epslon}
  \mathcal{D}_{\mu} \epsilon^i = \big(\partial_\mu - \tfrac{1}{4}
    \omega_\mu{}^{cd} \, \gamma_{cd} + \tfrac1{2} \, b_\mu +
    \tfrac{1}{2}\mathrm{i} \, A_\mu  \big) \epsilon^i + \tfrac1{2} \,
  \mathcal{V}_{\mu}{}^i{}_j \, \epsilon^j  \,.
\end{equation}

The covariant curvatures of the various gauge symmetries take the
following form,
\begin{align}
  \label{eq:curvatures}
  R(P)_{\mu \nu}{}^a  = & \, 2 \, \partial_{[\mu} \, e_{\nu]}{}^a + 2 \,
  b_{[\mu} \, e_{\nu]}{}^a -2 \, \omega_{[\mu}{}^{ab} \, e_{\nu]b} -
  \tfrac1{2} ( \bar\psi_{[\mu}{}^i \gamma^a \psi_{\nu]i} +
  \mbox{h.c.} ) \, , \nonumber\\[.2ex]
  R(Q)_{\mu \nu}{}^i = & \, 2 \, \mathcal{D}_{[\mu} \psi_{\nu]}{}^i -
  \gamma_{[\mu}   \phi_{\nu]}{}^i - \tfrac{1}{8} \, T^{abij} \,
  \gamma_{ab} \, \gamma_{[\mu} \psi_{\nu]j} \, , \nonumber\\[.2ex]
  R(A)_{\mu \nu} = & \, 2 \, \partial_{[\mu} A_{\nu ]} - \mathrm{i}
  \left( \tfrac12
    \bar{\psi}_{[\mu}{}^i \phi_{\nu]i} + \tfrac{3}{4} \bar{\psi}_{[\mu}{}^i
    \gamma_{\nu ]} \chi_i - \mbox{h.c.} \right) \, , \nonumber\\[.2ex]
  R(\mathcal{V})_{\mu \nu}{}^i{}_j =& \, 2\, \partial_{[\mu}
  \mathcal{V}_{\nu]}{}^i{}_j +
  \mathcal{V}_{[\mu}{}^i{}_k \, \mathcal{V}_{\nu]}{}^k{}_j  +  2 (
    \bar{\psi}_{[\mu}{}^i \, \phi_{\nu]j} - \bar{\psi}_{[\mu j} \,
    \phi_{\nu]}{}^i )
  -3 ( \bar{\psi}_{[\mu}{}^i \gamma_{\nu]} \chi_j -
    \bar{\psi}_{[\mu j} \gamma_{\nu]} \chi^i ) \nonumber\\
& \, - \delta_j{}^i ( \bar{\psi}_{[\mu}{}^k \, \phi_{\nu]k} -
  \bar{\psi}_{[\mu k} \, \phi_{\nu]}{}^k )
  + \tfrac{3}{2}\delta_j{}^i (\bar{\psi}_{[\mu}{}^k \gamma_{\nu]}
  \chi_k - \bar{\psi}_{[\mu k} \gamma_{\nu]} \chi^k)  \, , \nonumber\\[.2ex]
  R(M)_{\mu \nu}{}^{ab} = & \,
  \, 2 \,\partial_{[\mu} \omega_{\nu]}{}^{ab} - 2\, \omega_{[\mu}{}^{ac}
  \omega_{\nu]c}{}^b
  - 4 f_{[\mu}{}^{[a} e_{\nu]}{}^{b]}
  + \tfrac12 (\bar{\psi}_{[\mu}{}^i \, \gamma^{ab} \,
  \phi_{\nu]i} + \mbox{h.c.} ) \nonumber\\
& \, + ( \tfrac14 \bar{\psi}_{\mu}{}^i   \,
  \psi_{\nu}{}^j  \, T^{ab}{}_{ij}
  - \tfrac{3}{4} \bar{\psi}_{[\mu}{}^i \, \gamma_{\nu]} \, \gamma^{ab}
  \chi_i
  - \bar{\psi}_{[\mu}{}^i \, \gamma_{\nu]} \,R(Q)^{ab}{}_i
  + \mbox{h.c.} ) \, , \nonumber\\[.2ex]
  R(D)_{\mu \nu} = & \,2\,\partial_{[\mu} b_{\nu]} - 2 f_{[\mu}{}^a
  e_{\nu]a}
  - \tfrac{1}{2} \bar{\psi}_{[\mu}{}^i \phi_{\nu]i} + \tfrac{3}{4}
    \bar{\psi}_{[\mu}{}^i \gamma_{\nu]} \chi_i
    - \tfrac{1}{2} \bar{\psi}_{[\mu i} \phi_{\nu]}{}^i + \tfrac{3}{4}
  \bar{\psi}_{[\mu i} \gamma_{\nu]} \chi^i \,,  \nonumber\\[.2ex]
  R(S)_{\mu\nu}{}^i  = \,&  2\,{\cal D}_{[\mu}\phi_{\nu]}{}^i
  -2 f_{[\mu}{}^a\gamma_a\psi_{\nu]}{}^i
  -\ft18 \Slash{D} T_{ab}{}^{ij}\gamma^{ab}\gamma_{[\mu} \psi_{\nu]\, j}
     -\tfrac32 \gamma_a\psi_{[\mu}{}^i\,\bar\psi_{\nu]}{}^j\gamma^a{\chi}_j
     \nonumber\\
     \,& +\ft14 R({\cal V})_{ab}{}^i{}_j\gamma^{ab}
     \gamma_{[\mu}\psi_{\nu]}{}^j
     +\ft12 \mathrm{i}
     R(A)_{ab}\gamma^{ab}\gamma_{[\mu}\psi_{\nu]}{}^i
     \,,\nonumber\\[.2ex]
     R(K)_{\mu\nu}{}^a = \,& 2 \,{\cal D}_{[\mu} f_{\nu]}{}^a
     -\ft14\big(\bar{\phi}_{[\mu}{}^i\gamma^a\phi_{\nu]i}
     +\bar{\phi}_{[\mu i} \gamma^a\phi_{\nu]}{}^i\big)  \nonumber\\
     &\,
     +\tfrac14\big(\bar{\psi}_{\mu }{}^iD_b T^{ba}{}_{ij}\psi_{\nu}{}^j
     -3\, e_{[\mu}{}^a\psi_{\nu]}{}^i\Slash{D}\chi_i +\ft32
     D\,\bar{\psi}_{[\mu}{}^i\gamma^a\psi_{\nu]j}
     -4\,\bar{\psi}_{[\mu}{}^i\gamma_{\nu]}D_b R(Q)^{ba}{}_i
     +\mbox{h.c.}\big)\,. \nonumber\\
     &{~}
\end{align}
There are three conventional constraints (which have already been
incorporated in \eqref{eq:curvatures},
\begin{align}
  \label{eq:conv-constraints}
  &R(P)_{\mu \nu}{}^a =  0 \, , \nonumber \\[1mm]
  &\gamma^\mu R(Q)_{\mu \nu}{}^i + \tfrac32 \gamma_{\nu}
  \chi^i = 0 \, , \nonumber\\[1mm]
  &
  e^{\nu}{}_b \,R(M)_{\mu \nu a}{}^b - \mathrm{i} \tilde{R}(A)_{\mu a} +
  \tfrac1{8} T_{abij} T_\mu{}^{bij} -\tfrac{3}{2} D \,e_{\mu a} = 0
  \,,
\end{align}
which are S-supersymmetry invariant. They determine the fields
$\omega_{\mu}{}^{ab}$, $\phi_\mu{}^i$ and $f_{\mu}{}^a$ as follows,
\begin{align}
  \label{eq:dependent}
  \omega_\mu^{ab} =&\, -2e^{\nu[a}\partial_{[\mu}e_{\nu]}{}^{b]}
     -e^{\nu[a}e^{b]\sigma}e_{\mu c}\partial_\sigma e_\nu{}^c
     -2e_\mu{}^{[a}e^{b]\nu}b_\nu   \nonumber\\
      &\, -\ft{1}{4}(2\bar{\psi}_\mu^i\gamma^{[a}\psi_i^{b]}
     +\bar{\psi}^{ai}\gamma_\mu\psi^b_i+{\rm h.c.}) \,,\nonumber\\
     \phi_\mu{}^i  =& \, \tfrac12 \left( \gamma^{\rho \sigma} \gamma_\mu -
    \tfrac{1}{3} \gamma_\mu \gamma^{\rho \sigma} \right) \left(
    \mathcal{D}_\rho
    \psi_\sigma{}^i - \tfrac{1}{16} T^{abij} \gamma_{ab} \gamma_\rho
    \psi_{\sigma j} + \tfrac{1}{4} \gamma_{\rho \sigma} \chi^i \right)
    \,,  \nonumber\\
    f_\mu{}^{\mu}  =& \, \tfrac{1}{6} R(\omega,e) - D - \left(
      \tfrac1{12} e^{-1}
    \varepsilon^{\mu \nu \rho \sigma} \bar{\psi}_\mu{}^i \, \gamma_\nu
    \mathcal{D}_\rho \psi_{\sigma i} - \tfrac1{12} \bar{\psi}_\mu{}^i
    \psi_\nu{}^j T^{\mu \nu}{}_{ij} - \tfrac1{4} \bar{\psi}_\mu{}^i
    \gamma^\mu \chi_i +
    \mbox{h.c.} \right) \, .
\end{align}
We will also need the bosonic part of the expression for the
uncontracted connection $f_\mu{}^a$,
\begin{equation}
  \label{eq:f-bos-uncon}
  f_\mu{}^a= \tfrac12 R(\omega,e)_\mu{}^a -\tfrac14 \big(D+\tfrac13
  R(\omega,e)\big) e_\mu{}^a -\tfrac12\mathrm{i}\tilde R(A)_\mu{}^a +
  \tfrac1{16} T_{\mu b} {}^{ij} T^{ab}{}_{ij} \,,
\end{equation}
where $R(\omega,e)_\mu{}^a= R(\omega)_{\mu\nu}{}^{ab} e_b{}^\nu$ is
the non-symmetric Ricci tensor, and $R(\omega,e)$ the corresponding
Ricci scalar. The curvature $R(\omega)_{\mu\nu}{}^{ab}$ is associated
with the spin connection field $\omega_\mu{}^{ab}$, given in
(\ref{eq:dependent}).

The transformations of $\omega_{\mu}{}^{ab}$, $\phi_\mu{}^i$ and
$f_{\mu}{}^a$ are induced by the constraints
\eqref{eq:conv-constraints}. We present their Q- and S-supersymmetry
variations, as well as the transformations under conformal boosts,
below,
\begin{align}
  \label{eq:dep-variations}
  \delta\omega_\mu{}^{ab} =&\,-\tfrac12 \bar \epsilon^i\gamma^{ab}
  \phi_{\mu i} -\tfrac12 \bar\epsilon^i\psi_\mu{}^j \,T^{ab}{}_{ij} +
  \tfrac34 \bar\epsilon^i\gamma_\mu\gamma^{ab} \chi_i  \nonumber\\
  &\, +\bar \epsilon^i\gamma_\mu \,R^{ab}{}_i(Q)- \tfrac12\bar
  \eta^i\gamma^{ab} \psi_{\mu i} + \mathrm{h.c.} +
  2\,\Lambda_\mathrm{K}{}^{[a} e_\mu{}^{b]}\,, \nonumber \\
  \delta\phi_\mu{}^i =&\, - 2\,f_\mu{}^a\gamma_a\epsilon^i + \ft14
  R({\cal V})_{ab}{}^{\, i}{}_{\!j}  \gamma^{ab}\gamma_\mu \epsilon^j
  + \tfrac1{2}\mathrm{i}  R(A)_{ab}\gamma^{ab}\gamma_\mu \epsilon^i
  -\tfrac1{8} \Slash{D} T^{ab\,ij} \gamma_{ab} \gamma_\mu \epsilon_j
  \nonumber\\
  &\, + \tfrac32 [(\bar \chi_j\gamma^a\epsilon^j)\gamma_a\psi_\mu{}^i
  -(\bar \chi_j\gamma^a\psi_\mu{}^j) \gamma_a \epsilon^i ]
  + 2\,\mathcal {D}_\mu\eta^i+
  \Lambda_\mathrm{K}{}^a\gamma_a\psi_\mu{}^i \,,\nonumber\\
  \delta f_\mu{}^a =&\, -\tfrac12 \bar\epsilon^i \psi_\mu{}^i\, D_b
  T^{ba}{}_{ij} -\tfrac34 e_\mu{}^a \bar \epsilon^i \Slash{D} \chi_i -
  \tfrac34 \bar\epsilon^i \gamma^a\psi_{\mu i}\,D \nonumber\\
  &\, +\bar\epsilon^i\gamma_\mu \,D_bR^{ba}{}_i(Q) +\tfrac12 \bar\eta^i
    \gamma^a\phi_{\mu i} + \mathrm{h.c.} +\mathcal{D}_\mu
    \Lambda_\mathrm{K}{}^a\,.
\end{align}
The transformations under S-supersymmetry and conformal boosts reflect
the structure of the underlying $\mathrm{SU}(2,2\vert 2)$ gauge
algebra. The presence of curvature constraints and of the non-gauge
fields $T_{abij}$, $\chi^i$ and $D$ induce deformations of the
Q-supersymmetry algebra, as is manifest in the above results, in
particular in \eqref{eq:curvatures} and \eqref{eq:dep-variations}.

Combining the conventional constraints \eqref{eq:conv-constraints}
with the various Bianchi identities one derives that not all the
curvatures are independent. For instance,
\begin{align}
  \label{eq:K-Bianchi}
  \varepsilon^{abcd} D_b \mathcal{R}(M)_{cd}{}^{ef} =&\, 2\,
  \varepsilon^{abc[e} \, R(K)_{bc}{}^{f]}+\ft92\, \eta^{a[e}\bar{\chi}^i
  \gamma^{f]}\chi_i\nonumber\\
  &\,
  +\ft12 \big[3 \bar{\chi}^i\gamma^a R(Q)^{efi}+\ft18
  D^b (T^{ab}{}_{ij}T^{efij})-\text{h.c.}\big]\,.
\end{align}

Furthermore it is convenient to modify two of the curvatures by
including suitable covariant terms,
\begin{align}
  \label{eq:mod-curv}
  {\cal R}(M)_{ab}{}^{\!cd} =\, &    R(M)_{ab}\,^{cd} + \ft1{16}\big(
  T_{abij}\,T^{cdij} + T_{ab}{}^{ij}\, T^{cd}{}_{ij}  \big)\,, \nonumber \\
  {\cal R}(S)_{ab}{}^i =\, & R(S)_{ab}{}^i + \ft34
  T_{ab}{}^{ij} \chi_j\,.
\end{align}
where we observe that $\gamma^{ab} \big(\mathcal{R}(S)-
R(S)\big)_{ab}{}^i=0$.  The modified curvature
$\mathcal{R}(M)_{ab}{}^{\!cd}$ satisfies the following relations,
\begin{align}
   \label{eq:RM-constraints}
  {\cal R}(M)_{\mu\nu}{}^{\!ab} \,e^\nu{}_b =\,& \mathrm{i} \tilde
  R(A)_{\mu\nu} e^{\nu a} +\ft32 D\,e_\mu{}^a\,, \nonumber\\
  \ft14 \varepsilon_{ab}{}^{ef} \,\varepsilon^{cd}{}_{gh} \,{\cal
    R}(M)_{ef}{}^{\!gh} =\,& {\cal R}(M)_{ab}{}^{\!cd} \,,\nonumber\\
  \varepsilon_{cdea}\,{\cal R}(M)^{cd\,e}{}_{\!b} =\,&
  \varepsilon_{becd} \,{\cal R}(M)_a{}^{\!e\,cd}  = 2\tilde{R}(D)_{ab} =
  2\mathrm{i} R(A)_{ab}\,.
\end{align}
The first of these relations corresponds to the third constraint given
in~\eqref{eq:conv-constraints}, while the remaining equations follow
from combining the curvature constraints with the Bianchi
identities. Note that the modified curvature does not satisfy the pair
exchange property; instead we have,
\begin{equation}
{\cal R}(M)_{ab}{}^{\!cd} = {\cal R}(M)^{cd}{}_{\!ab} + 4 \mathrm{i}
\delta^{[c}_{[a} \, \tilde R(A) _{b]}^{~}{}^{\!d]}\,.
\end{equation}

We now turn to the fermionic constraint given in
\eqref{eq:conv-constraints} and its consequences for the modified
curvature defined in \eqref{eq:mod-curv}. First we note that the
constraint on $R(Q)_{\mu\nu}{}^i$ implies that this curvature is
anti-selfdual, as follows from contracting the constraint with
$\gamma^\nu\,\gamma_{ab}$,
\begin{equation}
  \label{eq:selfdualRQ2}
  \tilde R(Q)_{\mu\nu}{}^i= - R(Q)_{\mu\nu}{}^i\,.
\end{equation}
Furthermore, combination of the Bianchi identity and the constraint on
$R(Q)_{\mu\nu i}$ yields the following condition on the modified
curvature $\mathcal{R}(S)_{ab}{}^i$,
\begin{equation}
  \label{eq:Bianchi-RS}
  \gamma^a\tilde {\cal R}(S)_{ab}{}^i = 2\, D^a\tilde R(Q)_{ab}{}^i =
  -2\,D^aR(Q)_{ab}{}^i \,.
\end{equation}
This identity (upon contraction with $\gamma^b\gamma_{cd}$) leads to
the following identity on the anti-selfdual part of
$\mathcal{R}(S)_{ab}{}^i$,
\begin{equation}
  {\cal R}(S)_{ab}{}^i-\tilde{\cal R}(S)_{ab}{}^i =
  2\,\Slash{D}\big( R(Q)_{ab}{}^i + \ft34 \gamma_{ab} \chi^i\big)\,.
\end{equation}

Finally we note the following useful identities for products of
(anti)selfdual tensors,
\begin{align}
  \label{eq:dual-tensor-products}
  G^\pm_{[a[c}\, H_{d]b]}^\pm =&\, \pm\ft18
  G^\pm_{ef} \,H^{\pm ef} \,\varepsilon_{abcd} -\ft14( G^\pm_{ab}\,
  H^\pm_{cd} +G^\pm_{cd}\, H^\pm_{ab}) \,,\nonumber \\
  G^{\pm}_{ab} \,H^{\mp cd} + G^{\pm cd} \, H^{\mp}_{ab} =&\, 4 \delta^{[c}_{[a}
  G^{\pm}_{b]e} \, H^{\mp d]e} \,,\nonumber \\
  \ft12 \varepsilon^{abcd} \,G^{\pm}_{[c}{}^e \, H^\pm_{d]e} =&\, \pm
  G^{\pm [a}{}_{e} \, H^{\pm b]e}\,,\nonumber \\
  G^{\pm ac}\,H^\pm_c{}^b + G^{\pm bc}\,H^\pm_c{}^a=&\, -\ft12
  \eta^{ab}\, G^{\pm cd}\,H^\pm_{cd} \,,\nonumber \\
  G^{\pm ac}\,H^\mp_c{}^b =&\, G^{\pm bc}\,H^\mp_c{}^a\,, \qquad
  G^{\pm ab}\,H^\mp_{ab} =0 \,.
\end{align}

\section{Covariantization under special superconformal boosts}
\label{App:cov-supercon-boosts}
\setcounter{equation}{0}
In principle covariant (multiple) derivatives are
defined by the standard procedure by adding gauge fields to absorb all
symmetry variations proportional to derivatives of the transformation
parameters. In this procedure the gauge field $f_\mu{}^a$ associated
with the conformal boosts (parametrized by $\Lambda_\mathrm{K}{}^a$)
appears somewhat indirectly, because the only other fields that
transform under the conformal boosts are the gauge fields $b_\mu$,
$\omega_\mu{}^{ab}$ and $\phi_\mu{}^i$. Therefore supercovariant
derivatives of fields that are themselves invariant, will transform
under these K-transformations, and usually these variations take a
relatively simple form. We give some examples for a scalar field
$\phi$, a spinor field $\psi$, and a tensor field $t_{ab}$, each of
Weyl weight $w$,
\begin{align}
  \label{eq:3K-var-D-field}
  \delta_\mathrm{K} D_a \phi= &\, -w\,\Lambda_{\mathrm{K}a}
  \phi\,,  \nonumber \\
  \delta_\mathrm{K}D_a t_{bc}=&\, - w\,\Lambda_{\mathrm{K}a} t_{bc}
   +2\, t_{a[b}  \Lambda_{\mathrm{K}c]}
   -2\, \eta_{a[b} t_{c]d} \,\Lambda_\mathrm{K}{}^d\,, \nonumber \\
  \delta_\mathrm{K} D_a \psi =&\, \big[ - w\,\Lambda_{\mathrm{K}a}
   +\tfrac12 \Lambda_{\mathrm{K}}{}^b \gamma_{ab}\big]\psi\,.
\end{align}
These transformation rules simplify for certain contractions, such as
in $D^at_{ab}$ or $\Slash{D}\psi$,
\begin{align}
  \label{eq:K-var-contr}
   \delta_\mathrm{K}D^a t_{ab}=&\,(2-w) \Lambda_\mathrm{K}{}^a t_{ab}
   \,,\nonumber \\
   \delta_\mathrm{K}D_{[a} t_{bc]}=&\,(2-w) \Lambda_{\mathrm{K}[a}
   t_{bc]}    \,,\nonumber \\
   \delta_\mathrm{K} \Slash{D} \psi =&\, (\tfrac32 -w)
   \Slash{\Lambda}_\mathrm{K} \,\psi\,,
\end{align}
showing, for instance, that the Dirac operator on a spinor field of
weight $w=\tfrac32$ is invariant.

Applying an extra covariant derivative we explicitly indicate the
presence of the K-connection field $f_\mu{}^a$,
\begin{align}
  \label{eq:K-double-der}
   D_\mu D_{a} \phi=&\, \mathcal{D}_\mu D_a\phi +wf_{\mu
     a} \,\phi  \,,\nonumber \\
   D_\mu D^a t_{ab}=&\,\mathcal{D}_\mu D^a t_{ab} +(w-2)
   f_\mu{}^a\,  t_{ab}    \,,\nonumber \\
   D_\mu\Slash{D} \psi =&\, \mathcal{D}_\mu \Slash{D} \psi
   +(w-\tfrac32) f_\mu{}^a \gamma_a\psi \,,
\end{align}
where $\mathcal{D}_\mu$ denotes the covariant derivative without
including the field $f_\mu{}^a$. Under K-transformations these
multiple derivatives transform as,
\begin{align}
  \label{eq:K-var-double-der}
   \delta_\mathrm{K} D_\mu D_{a} \phi=&\, -(w+1)
   \big[\Lambda_{\mathrm{K}\mu}  D_a+ \Lambda_{\mathrm{K}a}
   D_\mu\big] \phi + e_{\mu a} \Lambda_\mathrm{K}{}^b D_b \phi
   \,,\nonumber \\
   \delta_\mathrm{K} D_\mu D^a t_{ab}=&\,-(w+1) \Lambda_{\mathrm{K}\mu}
   D^at_{ab}-\Lambda_{\mathrm{K}b} D^at_{a\mu} + e_{\mu b}
   \Lambda_\mathrm{K}{}^c D^at_{ac}
   +(2-w) \Lambda_\mathrm{K}{}^a D_\mu t_{ab}   \,,\nonumber \\
   \delta_\mathrm{K} D_\mu\Slash{D} \psi =&\,
   \big[-(w+1)\Lambda_{\mathrm{K}\mu}
   +\tfrac12 \Lambda_\mathrm{K}{}^a\gamma_{\mu a}\big]
   \Slash{D} \psi +(\tfrac32-w) \Slash{\Lambda}_\mathrm{K} D_\mu \psi \,.
\end{align}
Contracting the first equation with $e^{a\mu}$ shows that the
conformal D'Alembertian transforms under K-transformations as
$\delta_\mathrm{K} \Box_\mathrm{c} \phi= -2(w-1)\Lambda_\mathrm{K}{}^a
D_a\phi$, which vanishes for $w=1$.

This pattern repeats itself when considering even higher
derivatives. We present the following results,
\begin{align}
  \label{eq:cubic-quartic-der}
  D_\mu\Box_\mathrm{c}\phi =&\,  \mathcal{D}_\mu\Box_\mathrm{c}\phi
  +2(w-1)f_\mu{}^a D_a \phi\,,  \nonumber\\
  \Box_\mathrm{c} \Box_\mathrm{c}\phi =&\,  \mathcal{D}_\mu D^\mu
  \Box_\mathrm{c}\phi +(w+2)f_{\mu}{}^\mu \Box_\mathrm{c} \phi + 2(w-1)
  f_{\mu a} D^\mu D^a\phi \,,   \nonumber \\
  \Box_\mathrm{c} \Slash{D}\psi =&\, \mathcal{D}_\mu D^\mu
  \Slash{D}\psi +\big[(w+1) f_\mu{}^\mu - \tfrac12 f_{\mu
    a}\gamma^{\mu a}\big]
  \Slash{D}\psi +(w-\tfrac32) f_{\mu a} \gamma^a D^\mu \psi\,,
\end{align}
and,
\begin{align}
  \label{eq:K-var-cubic-quartic}
   \delta_\mathrm{K} \Box_\mathrm{c} \Box_\mathrm{c} \phi=&\, -2(w-1)
   \Lambda_{\mathrm{K}}{}^a \Box_\mathrm{c} D_a\phi - 2(w+1)
   \Lambda_\mathrm{K}{}^a D_a \Box_\mathrm{c} \phi \nonumber\\
   =&\, -2w \Lambda_{\mathrm{K}}{}^a \big[\Box_\mathrm{c}D_a\phi +
   D_a \Box_\mathrm{c}\big]\phi
   +2\Lambda_\mathrm{K}{}^a\big[\Box_\mathrm{c}D_a
   -D_a\Box_\mathrm{c}\big]\phi\, \,,\nonumber \\
   \delta_\mathrm{K} \Box_\mathrm{c} \Slash{D} \psi =&\,
   -(2w-1)\Lambda_{\mathrm{K}}{}^a D_a\Slash{D}\psi -\tfrac12
   \Slash{\Lambda}_\mathrm{K} \big[(2w-1)\Box_\mathrm{c} +
   [\Slash{D},\Slash{D}] \big]\psi\,.
\end{align}
For future use we have evaluated the previous two variations for the
fields $A$ and $\Psi_i$, which have weights $w=0,\tfrac12$,
respectively. In this case all the terms cubic and quadratic in
derivatives in \eqref{eq:K-var-cubic-quartic} appear with a certain
degree of anti-symmetry, such that they become proportional to
curvatures. Upon substituting the results for the various curvatures,
one obtains \eqref{eq:K-var-A-Psi}.

\section{Multiplication of chiral multiplets}
\label{App:chiral-multiplets}
\setcounter{equation}{0}
In this appendix we summarize the product rules for two chiral
supermultiplets and the Taylor expansion for functions of these
multiplets. In the local supersymmetry setting, we will usually be
dealing with homogeneous functions of chiral multiplets with equal
Weyl weight so that a scaling weight under Weyl transformations can
be assigned to the function.

The product of two chiral multiplets, specified by the component
fields $\big(A,\Psi_i,B_{ij}, F^-_{ab}, \Lambda_i,C\big)$ and
$\big(a,\psi_i,b_{ij}, f^-_{ab}, \lambda_i,c\big)$, respectively,
leads to the following decomposition,
\begin{align}
  \label{eq:chiral-mult-prod}
  &\big(A,\Psi_i,B_{ij}, F^-_{ab}, \Lambda_i,C\big)\otimes
  \big(a,\psi_i,b_{ij}, f^-_{ab}, \lambda_i,c\big)= \nonumber\\
  &\quad
  \big(A\,a\,,\, A\,\psi_i+a\,\Psi_i,A\,b_{ij} + a\,B_{ij}
  - \bar\Psi_{(i}\psi_{j)}\,,  \nonumber\\
  &\quad
  A\,f^-_{ab} + a\,F^-_{ab} -\tfrac14 \varepsilon^{ij} \bar
  \Psi_i\gamma_{ab}\psi_j \,,\nonumber\\
  &\quad
  A\,\lambda_i +a\,\Lambda_i -\tfrac12\varepsilon^{kl} (B_{ik}\,\psi_l
  +b_{ik}\,\Psi_l) -\tfrac14(F^-_{ab}\gamma^{ab}\psi_i
  +f^-_{ab}\gamma^{ab}\Psi_i ) \,,\nonumber\\
  &\quad
  A\,c+a\,C-\tfrac 12 \varepsilon^{ik}\varepsilon^{jl} B_{ij}\,b_{kl} +
  F^-_{ab}\, f^{-ab} +\varepsilon^{ij} (\bar\Psi_i\lambda_j+
  \bar\psi_i\Lambda_j)\big) \;.
\end{align}

A function $\mathcal{G}(\Phi)$ of chiral superfields $\Phi^I$ defines
a chiral superfield, whose component fields take the following form,
\begin{align}
  \label{eq:chiral-mult-exp}
  A\vert_\mathcal{G} =&\, \mathcal{G}(A) \,,\nonumber\\
  \Psi_i\vert_\mathcal{G} =&\, \mathcal{G}(A)_I \,\Psi_i{}^I
  \,,\nonumber\\
  B_{ij}\vert_\mathcal{G} =&\, \mathcal{G}(A)_I\, B_{ij}{}^I -\tfrac12
  \mathcal{G}(A)_{IJ} \,\bar \Psi_{(i}{}^I
  \Psi_{j)}{}^J \,,\nonumber\\
  F_{ab}^-\vert_\mathcal{G} =&\, \mathcal{G}(A)_I \,F_{ab}^-{}^I -\tfrac18
  \mathcal{G}(A)_{IJ}\, \varepsilon^{ij} \bar
  \Psi_{i}{}^I \gamma_{ab} \Psi_{j}{}^J \,,\nonumber\\
  \Lambda_{i}\vert_\mathcal{G} =&\, \mathcal{G}(A)_I \,\Lambda_{i}{}^I
  -\tfrac12
  \mathcal{G}(A)_{IJ}\big[B_{ij}{}^I   \varepsilon^{jk} \Psi_{k}{}^J
   +\tfrac12 F^{-}_{ab}{}^I\gamma^{ab} \Psi_{k}{}^J\big] \nonumber\\
   &\,
   + \tfrac1{48}  \mathcal{G}(A)_{IJK}\,\gamma^{ab} \Psi_i{}^I \,
   \varepsilon^{jk} \bar
   \Psi_{j}{}^J \gamma_{ab}  \Psi_{k}{}^K \,,\nonumber\\
   C\vert_\mathcal{G} =&\, \mathcal{G}(A)_I\, C^I  -\tfrac14
   \mathcal{G}(A)_{IJ}\big[ B_{ij}{}^I B_{kl}{}^J\,
   \varepsilon^{ik} \varepsilon^{jl}
   -2\, F^{-}_{ab}{}^I F^{-abJ} +4\,\varepsilon^{ik} \bar
   \Lambda_i{}^I \Psi_j{}^J\big]  \,,\nonumber\\
        &\,   +\tfrac14 \mathcal{G}(A)_{IJK} \big[ \varepsilon^{ik}
        \varepsilon^{jl}
        B_{ij}{}^I \Psi_{k}{}^J \Psi_{l}{}^K  -\tfrac12
        \varepsilon^{kl} \bar\Psi_{k}{}^I F^{-}_{ab}{}^J\gamma^{ab}
        \Psi_{l}{}^K\big] \nonumber\\
        &\,
        + \tfrac1{192} \mathcal{G}(A)_{IJKL} \,\varepsilon^{ij}  \bar
        \Psi_{i}{}^I \gamma_{ab} \Psi_{j}{}^J \,\varepsilon^{kl}  \bar
        \Psi_{k}{}^K \gamma_{ab} \Psi_{l}{}^L\,.
\end{align}
This result follows straightforwardly from expanding the
superfield expression in powers of the fermionic coordinates.

\section{Reduced chiral multiplets}
\label{App:reduced-multiplets}
\setcounter{equation}{0}
Chiral multiplets can be consistently reduced by imposing a reality
constraint. This usually requires specific values for the Weyl and
chiral weights. The two cases that are relevant are the vector
multiplet, which arises upon reduction from a scalar chiral multiplet,
and the Weyl multiplet, which is a reduced anti-selfdual chiral tensor
multiplet. Both reduced multiplets require weight $w=1$.

We will denote the components of the $w=1$ multiplet that describes
the vector multiplet by
$(A,\Psi,B,F^-,\Lambda,C)\vert_{\text{vector}}$. The constraint for a
scalar chiral supermultiplet reads, $\varepsilon^{ij}\,\bar
D_i\gamma_{ab}D_j \Phi = [\varepsilon^{ij}\,\bar D_i\gamma_{ab}D_j
\Phi ]^\ast$, which implies that
$C\vert_{\text{vector}}$ and $\Lambda_i\vert_{\text{vector}}$ are
expressed in terms of the lower components of the multiplet, and
imposes a reality constraint on $B\vert_{\text{vector}}$ and a Bianchi
identity on $F^-\vert_{\text{vector}}$
\cite{Firth:1974st,deRoo:1980mm,deWit:1980tn}. The latter implies that
$F^-\vert_{\text{vector}}$ can be expressed in terms of a gauge field
$W_{\mu}$. This feature is not affected by the presence of the
superconformal background field.

Denoting the independent components of the vector multiplet by
$(X,\Omega,Y,F^-)$, the identification with the chiral multiplet
components is as follows,
\begin{align}
  \label{eq:vect-mult}
  A\vert_{\text{vector}}=&\,X\,,\nonumber\\
  \Psi_i\vert_{\text{vector}}=&\, \Omega_i\,,\nonumber\\
  B_{ij}\vert_{\text{vector}}=&\, Y_{ij}
  =\varepsilon_{ik}\varepsilon_{jl}Y^{kl}\,,\nonumber\\
  F_{ab}^-\vert_{\text{vector}}=&   \big(\delta_{ab}{}^{cd} -\tfrac12
    \varepsilon_{ab}{}^{cd}\big) e_c{}^\mu e_d{}^\nu \,\partial_{[\mu}
    W_{\nu]}\nonumber\\
    &\,
  +\tfrac14\big[\bar{\psi}_{\rho}{}^i\gamma_{ab} \gamma^\rho\Omega^{j}
  + \bar{X}\,\bar{\psi}_\rho{}^i\gamma^{\rho\sigma}\gamma_{ab}
  \psi_\sigma{}^j
  - \bar{X}\, T_{ab}{}^{ij}\big]\varepsilon_{ij}  \,,\nonumber\\
  \Lambda_i\vert_{\text{vector}}
  =&\,-\varepsilon_{ij}\Slash{D}\Omega^j\nonumber\\
  C\vert_{\text{vector}}= &\,-2\, \Box_\mathrm{c}  \bar X  -\tfrac14  F_{ab}^+\,
   T^{ab}{}_{ij} \varepsilon^{ij} - 3\,\bar\chi_i \Omega^i\,.
\end{align}
The Bianchi identity on $F_{ab}$ can be written as,
\begin{align}
\label{eq:Bianchi-vector} D^b\left(F_{ab}^+-F_{ab}^- +\ft14 X
  T_{abij}\varepsilon^{ij}-\ft14 \bar{X}
  T_{ab}{}^{ij}\varepsilon_{ij}\right)
  +\ft34\left(\bar{\chi}_i\gamma_a \Omega_j
  \varepsilon^{ij}-\bar{\chi}^i\gamma_a\Omega^j
  \varepsilon_{ij}\right)=0&\,,
\end{align}
and the reality constraint on $Y_{ij}$ is included
in~\eqref{eq:vect-mult}.

The Q- and S-supersymmetry transformations for the vector multiplet
take the form,
\begin{align}
  \label{eq:variations-vect-mult}
  \delta X =&\, \bar{\epsilon}^i\Omega_i \,,\nonumber\\
  \delta\Omega_i =&\, 2 \Slash{D} X\epsilon_i
     +\ft12 \varepsilon_{ij}  F_{\mu\nu}
   \gamma^{\mu\nu}\epsilon^j +Y_{ij} \epsilon^j
     +2X\eta_i\,,\nonumber\\
  \delta W_{\mu} = &\, \varepsilon^{ij} \bar{\epsilon}_i
  (\gamma_{\mu} \Omega_j+2\,\psi_{\mu j} X)
  + \varepsilon_{ij}
  \bar{\epsilon}^i (\gamma_{\mu} \Omega^{j} +2\,\psi_\mu{}^j
  \bar X)\,,\nonumber\\
\delta Y_{ij}  = &\, 2\, \bar{\epsilon}_{(i}
  \Slash{D}\Omega_{j)} + 2\, \varepsilon_{ik}
  \varepsilon_{jl}\, \bar{\epsilon}^{(k} \Slash{D}\Omega^{l)
  } \,,
\end{align}
and, for $w=1$, are in clear correspondence with the supersymmetry
transformations of generic scalar chiral multiplets given in
\eqref{eq:conformal-chiral}.

Subsequently we turn to the Weyl multiplet, which is a chiral
anti-selfdual tensor multiplet subject to $\bar D_i\gamma^{ab}D_j
\,\Phi_{ab}{}^{ij} = [\bar D_i\gamma^{ab}D_j \,\Phi_{ab}{}^{ij}
]^\ast$. Its chiral superfield components take the following form,
\begin{align}
  \label{eq:W-mult}
  A_{ab}\vert_{W}   =&\,T_{ab}{}^{ij}\varepsilon_{ij}\,,\nonumber \\
  \Psi_{abi}\vert_{W} =&\, 8\, \varepsilon_{ij}R(Q)^j_{ab} \,,\nonumber\\
  B_{abij}\vert_{W}  =&\, -8 \,\varepsilon_{k(i}R({\cal V})_{ab}^-{}^k{}_{j)} \,,\nonumber\\
  \left(F^{-}_{ab}\right){}^{cd}\vert_{W}  =&\, -8 \,{\cal R}(M)_{ab}^-{}^{\!cd} \,,\nonumber\\
  \Lambda_{abi}\vert_{W} =&\, 8\left(\mathcal{R}(S)_{abi}^- + \ft34 \gamma_{ab}\Slash{D}\chi_i\right)
  \,,\nonumber\\
  C_{ab}\vert_{W} =&\,  4 D_{[a} \,D^cT_{b]c\,ij}
  \varepsilon^{ij}-\text{dual} \,.
\end{align}
We give the Q- and S-supersymmetry variations for the first few
components,
\begin{align}
  \label{eq:variations-W-mult}
    \delta T_{ab}{}^{ij} =&\, 8 \,\bar{\epsilon}^{[i} R(Q)_{ab}{}^{j]} \,
  , \nonumber \\[.2ex]
  \delta R(Q)_{ab}{}^i =&\, - \tfrac12 \Slash{D}T_{ab}{}^{ij}
  \,\epsilon_j +   R({\cal V})^-{}_{\!\!ab}{}^i{}_j \, \epsilon^j
    - \tfrac12 {\cal R}(M)_{ab}{}^{\!cd}\, \gamma_{cd} \epsilon^i +\tfrac18
    T_{cd}{}^{ij}\,\gamma^{cd}  \gamma_{ab} \,\eta_j  \,,\nonumber\\[.2ex]
    \delta R(\mathcal{V})^-{}_{\!\!ab}{}^i{}_j
  =&\,2\bar\epsilon_j\Slash{D}R(Q)_{ab}{}^i -2\bar\epsilon^i\big(
  \mathcal{R}(S)_{abj}^- +\tfrac34
  \gamma_{ab}\Slash{D}\chi_j\big) \nonumber\\
  &\,
  +\bar\eta_j(2R(Q)_{ab}{}^i+3\gamma_{ab}\chi^i) -
  (\mathrm{traceless})\,,  \nonumber\\[.2ex]
  \delta \mathcal{R}(M)_{ab}^-{}^{\!cd}=&\,\ft12 \bar{\epsilon}_i
  \Slash{D}\gamma^{cd}R(Q)_{ab}{}^i-\ft12
  \bar{\epsilon}^i\gamma^{cd}\big(\mathcal{R}(S)^-_{abi}+\ft34
  \gamma_{ab}\Slash{D}\chi_i\big)\nonumber\\[.2ex]
  &\,-\bar{\eta}_i\gamma_{ab} R(Q)^{cdi}-\ft12 \bar{\eta}_i
  \gamma^{cd} R(Q)_{ab}{}^i-\ft34
  \bar{\eta}_i\gamma_{ab}\gamma^{cd}\chi^i\,.
\end{align}

A scalar chiral multiplet with $w=2$ is obtained by squaring the Weyl
multiplet. The various scalar chiral multiplet components are given by,
\begin{align}
  \label{eq:W-squared}
  A\vert_{W^2}   =&\,(T_{ab}{}^{ij}\varepsilon_{ij})^2\,,\nonumber \\[.2ex]
  \Psi_i\vert_{W^2} =&\, 16\, \varepsilon_{ij}R(Q)^j_{ab} \,T^{klab}
  \, \varepsilon_{kl} \,,\nonumber\\[.2ex]
  B_{ij}\vert_{W^2}  =&\, -16 \,\varepsilon_{k(i}R({\cal
    V})^k{}_{j)ab} \, T^{lmab}\,\varepsilon_{lm} -64
  \,\varepsilon_{ik}\varepsilon_{jl}\,\bar R(Q)_{ab}{}^k\, R(Q)^{l\,ab}
  \,,\nonumber\\[.2ex]
  F^{-ab}\vert_{W^2}  =&\, -16 \,{\cal R}(M)_{cd}{}^{\!ab} \,
  T^{klcd}\,\varepsilon_{kl}  -16 \,\varepsilon_{ij}\, \bar
  R(Q)^i_{cd}  \gamma^{ab} R(Q)^{cd\,j}  \,,\nonumber\\[.2ex]
  \Lambda_i\vert_{W^2} =&\, 32\, \varepsilon_{ij} \,\gamma^{ab} R(Q)_{cd}^j\,
  {\cal R}(M)^{cd}{}_{\!ab}
  +16\,({\cal R}(S)_{ab\,i} +3 \gamma_{[a} D_{b]}  \chi_i) \,
  T^{klab}\, \varepsilon_{kl} \nonumber\\
  &\, -64\, R({\cal V})_{ab}{}^{\!k}{}_i \,\varepsilon_{kl}\,R(Q)^{ab\,l}
  \,,\nonumber\\[.2ex]
  C\vert_{W^2} =&\,  64\, {\cal R}(M)^{-cd}{}_{\!ab}\, {\cal
    R}(M)^-_{cd}{}^{\!ab}  + 32\, R({\cal V})^{-ab\,k}{}_l^{~} \,
  R({\cal V})^-_{ab}{}^{\!l}{}_k  \nonumber \\
  &\, - 32\, T^{ab\,ij} \, D_a \,D^cT_{cb\,ij} +
  128\,\bar{\mathcal{R}}(S)^{ab}{}_i  \,R(Q)_{ab}{}^i  +384 \,\bar
  R(Q)^{ab\,i} \gamma_aD_b\chi_i   \,.
\end{align}
These components can straightforwardly be substituted in the expression
for the higher-derivative couplings.

\end{appendix}

\providecommand{\href}[2]{#2}

\end{document}